\RequirePackage{etoolbox}
\csdef{input@path}{%
 {sty/}% cls, sty files
 {img/}% eps files
}%
\csgdef{bibdir}{bib/}% bst, bib files

\documentclass[ba]{imsart}
\pubyear{0000}
\volume{00}
\issue{0}
\doi{0000}
\firstpage{1}
\lastpage{1}

\usepackage{amsfonts}
\usepackage{amsthm}
\usepackage{amsmath}
\usepackage{natbib}
\usepackage[colorlinks,citecolor=blue,urlcolor=blue,filecolor=blue,backref=page]{hyperref}
\usepackage{graphicx}
\usepackage{wasysym}
\usepackage{algorithmic, algorithm}
\usepackage{url}

\startlocaldefs
\numberwithin{equation}{section}
\theoremstyle{plain}

\endlocaldefs

\def\var{\mbox{var}}
\def\cov{\mbox{cov}}

\def\bra{\langle}
\def\ket{\rangle}
\def\N{\mathcal N}
\def\R{\mathbb R}

\begin{document}

\begin{frontmatter}
\title{Calibration procedures for approximate Bayesian credible sets}%
\runtitle{Calibrating approximate credible sets}
\begin{aug}
\author{\fnms{Jeong Eun} \snm{Lee}\thanksref{addr1}\ead[label=e1]{kate.lee@auckland.ac.nz}},
\author{\fnms{Geoff K.} \snm{Nicholls}\thanksref{addr2,t1}\ead[label=e2]{nicholls@stats.ox.ac.uk}}
\and
\author{\fnms{Robin J.} \snm{Ryder}\thanksref{addr3}\ead[label=e3]{ryder@ceremade.dauphine.fr}}

\runauthor{J. E. Lee, G. K. Nicholls and R. J. Ryder}

\address[addr1]{
    The Department of Statistics,
    University of Auckland, Private Bag 92019,
    Auckland 1142,
    New Zealand
    \printead{e1}
}

\address[addr2]{
    Department of Statistics,
    24-29 St Giles,
    Oxford, OX1 3LG,
    UK
    \printead{e2}
}

\address[addr3]{
    CEREMADE, CNRS, Universit\'e Paris-Dauphine, PSL  University, 75016 Paris, France
    \printead{e3}
}

\thankstext{t1}{Corresponding author.}

\end{aug}

\begin{abstract}
We develop and apply
two calibration procedures for checking the coverage of approximate Bayesian credible sets including intervals estimated using Monte Carlo methods.
The user has an ideal prior and likelihood, but generates a credible set for an approximate posterior which is not proportional to the
product of ideal likelihood and prior. We estimate the realised posterior coverage achieved by the approximate credible set.
This is the coverage of the unknown ``true'' parameter if the data are a realisation of the user's ideal observation model conditioned on the parameter,
and the parameter is a draw from the user's ideal prior.
%In our setting the coverage is asymptotically exact when the user's exact prior
%and likelihood are used in an asymptotically exact Monte Carlo method, and otherwise it may be distorted.
In one approach we estimate the posterior coverage at the data by making a semi-parametric logistic regression
of binary coverage outcomes on simulated data against summary statistics evaluated on simulated data. In another we use Importance Sampling from the approximate posterior, windowing simulated data to fall close to the
observed data. %We give a Bayes Factor measuring the evidence for the operational posterior coverage to be below a user specified threshold.
We illustrate our methods on four examples.
\end{abstract}



\end{frontmatter}

\section{Introduction}

When we carry out Bayesian inference it is often convenient,
even when not strictly necessary, to make approximations. We work
with likelihoods and priors which only approximately equal those we would ideally use. Examples of popular approximations include Approximate Bayesian Computation \citep{pritchard99,marin2012approximate}, pseudo-likelihoods \citep{besag1975statistical}, synthetic likelihood \citep{wood2010statistical}, Variational Bayes \citep{jordan1999introduction} and Expectation Propagation \citep{minka2001expectation}.
%When we implement a Monte Carlo algorithm (for example, Metropolis Hastings MCMC) we may plug one of these approximations in as %an estimate of the posterior.
%The output of an approximate method is typically a sample from, or description of, an approximate posterior distribution.
If we use an approximate posterior distribution to get an approximate credible set with nominal level $\alpha$, we should expect the approximation to distort the coverage, so that the operational coverage of our approximate credible set is not $\alpha$.
In this paper we give a procedure for measuring the operational coverage. We ignore questions of goodness of fit. We are not aiming to calibrate coverage of the true parameter, but to measure the distortion in coverage due to target approximation. 
%We assume the ideal prior and likelihood
%are known and if we could compute with them, we would.

Our approach was inspired by \cite{geweke04}, and \cite{cook06} and later related papers
including \cite{fearnhead12}, \cite{prangle14} and \cite{yao18} which exploit an idea set out by \cite{monahan92}.
Let $\pi(\phi)$ be a prior for a parameter $\phi\in \Omega$, let $p(y|\phi)$ be an observation model for data $y\in \mathcal Y$
with
\[\pi(\phi|y)\propto \pi(\phi)p(y|\phi)\]
the posterior for $\phi$ given $y$.
Let $\tilde\pi(\theta)$ and $\tilde p(y|\theta)$ be an approximate prior and approximate likelihood for a parameter $\theta\in \Omega$ with
\[\tilde \pi(\theta|y)\propto \tilde \pi(\theta)\tilde p(y|\theta)\]
the approximate posterior for $\theta$ given $y$.
Suppose we simulate $\phi\sim \pi(\cdot)$, $y'\sim p(\cdot|\phi)$ and $\theta\sim
 \tilde \pi(\cdot|y')$.
%If $\tilde\pi(\theta|y)=\pi(\theta|y)$ (so, there is no approximation) then $\phi$ and $\theta$ are exchangeable
%and the marginal distribution $m(\theta)$ of $\theta$ is $m(\theta)=\pi(\theta)$, the prior.
The joint conditional distribution of $\phi$ and $\theta$,
$m(\theta,\phi|y')$ say, is
\begin{equation}\label{eq:iec}
m(\theta,\phi|y')=\pi(\phi|y')\tilde \pi(\theta|y'),
\end{equation}
so, conditional on $y'$, $\phi$ and $\theta$ are exchangeable if and only if there is no approximation and $\tilde\pi(\theta|y')=\pi(\theta|y')$ for all $\theta\in\Omega$.
The joint marginal distribution $m(\theta,\phi)$ is
\begin{equation}\label{eq:iem}
m(\theta,\phi)=\int \pi(\phi|y')\tilde\pi(\theta|y') p(y') dy'
\end{equation}
where
$
p(y')=\int_\Omega \pi(\phi)p(y'|\phi) d\phi
$
is the exact marginal likelihood, so $\phi$ and $\theta$ are marginally exchangeable if there is no approximation.

In work to date, Equation~\ref{eq:iem} has been taken as a starting point, as it gives a necessary condition on correctly distributed samples $\theta$,
which can be tested to check an approximation is good. For example, for $i=1,\ldots,M$, simulate
$\phi_{(i)}\sim \pi(\cdot)$, $y_{(i)}\sim p(\cdot|\phi_{(i)})$ and $\theta_{(i)}\sim \tilde \pi(\cdot|y_{(i)})$; here $y_{(i)}\in \mathcal Y$ is a data set,
$\phi_{(i)},\theta_{(i)}\in \Omega$ are parameter vectors, and the realisation
$\theta_{(i)}\sim \tilde \pi(\cdot|y_{(i)})$ might for example be the last sample in a MCMC run.
The parameter vectors $\phi_{(1)},\ldots,\phi_{(M)}$ and $\theta_{(1)},\ldots,\theta_{(M)}$
can be compared by a non-parametric test such as a rank test, if the parameters are scalar, and
otherwise comparing single components. Under the null, the two sets have the same distribution.
If we reject the null, this is evidence for $\tilde \pi(\theta|y)\ne \pi(\theta|y)$.
%A test for $\phi\sim \theta$ marginally checks a necessary condition for $\pi(\phi|y)=\tilde\pi(\theta|y)$.
\cite{geweke04} and \cite{cook06} use ideas along these lines to check for
correct MCMC (implementation, convergence) sampling of $\theta\sim \tilde \pi(\cdot|y')$, since in their
case simulation of $\theta\sim \pi(\cdot|y')$ is possible using MCMC and they want to check it is working correctly.
\cite{yao18} and \cite{talts18} move from testing MCMC convergence to diagnosing poor approximation of priors and likelihoods.

It is well known that this sort of check for $\tilde\pi(\theta|y)\simeq\pi(\theta|y)$, based on testing for exchangeable marginal distributions,
can be fooled by an approximation which is far from the posterior.
In particular if $0\le a\le 1$ and for all $y'\in \mathcal Y$ our approximation is a linear combination of prior and posterior,
$\tilde\pi(\theta|y')=a\pi(\theta|y')+(1-a)\pi(\theta)$ then
\begin{eqnarray}
% \nonumber to remove numbering (before each equation)
  m(\theta,\phi)&=&\int \pi(\phi|y')[a\pi(\theta|y')+(1-a)\pi(\theta)] p(y') dy' \nonumber \\
   &=& a\int \pi(\phi|y')\pi(\theta|y')p(y')dy' + (1-a)\pi(\phi)\pi(\theta)\label{eq:iep}
\end{eqnarray}
and $\theta$ and $\phi$ are marginally exchangeable. The test passes with $\tilde\pi(\theta|y)\ne\pi(\theta|y)$ for any sample size $M$. The case where $a=0$ (the approximate posterior is the prior) is discussed in \cite{prangle14} for Approximate Bayesian Computation (ABC) where it cannot be ignored,
as this sort of error is a real possibility in ABC. They treat this issue by conditioning $(\phi,y')$ on $y'\in A$ for some set $A\subset \mathcal Y$. One of our algorithms (Importance Sampling in Algorithm~\ref{al:2}) uses the same idea. We explain the connection between the two approaches below Equation~\ref{eq:pranglestyle}.
The focus shifts in \cite{rodrigues18} from testing for good calibration to recalibration of approximate samples. This approach is discussed further in Section~\ref{sec:recal}.
In brief, \cite{rodrigues18} estimate a recalibration map and use it to map ABC-samples onto the data in an ABC regression adjustment. By contrast we extract the closely related coverage error map at the data, as it gives us the realised coverage we are achieving for an arbitrary nominal or intended coverage. In earlier work \cite{menendez14} give a procedure for correcting a credible interval to give the nominal frequentist coverage for a parameter $\phi$ where a consistent estimator $\bar \phi$ is available.
In \cite{yao18} and \cite{talts18} the approximation framework is unrestricted, however these authors take Equation~\ref{eq:iem} as their starting point. They are interested in identifying how badly and in what ways the approximate distribution
$\tilde\pi(\theta|y)$ differs from the exact distribution $\pi(\theta|y)$. They expect an approximation error, so there seems little point in testing for one, but characterising and visualising any shift in distribution is still useful.

We assume that the desired output of an analysis is a credible set, and that we have a method for estimating the credible set which involves making an approximation. Is the estimated approximate credible set good, in the following sense? In the original analysis, without an approximation, the credible set is designed to achieve the nominal coverage $\alpha$ for a parameter $\phi$ with two assumed properties, $\phi\sim \pi(\cdot)$ and $y\sim p(\cdot|\phi)$, that is $\phi$ is a draw from the prior and the data $y$ inform $\phi$ through the observation model. We estimate the coverage our approximation actually achieves for a parameter $\phi$ satisfying these two properties. Coverage is usually taken to mean coverage of the unknown true parameter. Our definition of coverage is equivalent if we assume that the prior $\pi(\phi)$ is the true generative process for the unknown parameter $\phi$ and the observation model $p(y|\phi)$ is similarly correct. This is a shift from basing a test on Equation~\ref{eq:iem} to basing a bias estimate on Equation~\ref{eq:iec}.
This definition is appropriate as we focus on measuring approximation bias not model mispecification error. In Section~\ref{sec:cal} we introduce regression and importance-sampling (IS) methods for the purpose. Although most of our examples treat credible intervals for a real scalar variable, this is not a restriction. Our simulation-based methods apply to measurable credible sets, where they can be conveniently computed and specified. In our final example we work with a credible set for a finite random variable.

%We may be concerned to check the posterior coverage is acceptable, for example at least $\alpha-\delta$, with $\delta>0$ specified.
%Suppose the user supplies $\alpha$ and $\delta$, and a prior, likelihood and data, and we return an approximate credible set. We report as follows:
%``Here is a credible set $\hat C_y$ with nominal coverage at level $\alpha$. We assume the parameter
%is a draw from your prior and the data you provided are a draw from your observation model conditioned on the parameter.
%However this credible set was estimated under an approximation.
%We ran a test to see if our coverage differs greatly from what we would have achieved if we had not made an approximation.
%The Bayes factor (or if you prefer, posterior probability) for acceptable coverage at the data you gave us is $X$''.
%The procedure is straightforward and illustrated in Section~\ref{sec:part}.

The methodology we describe may be computationally costly, since it may involve repeating the approximate inference procedure $M$ times with $M$ large. In some settings this would defeat the purpose of using an approximate scheme, since these are usually chosen to provide rapid answers. There are some mitigating factors. The runs can be processed in parallel, thus decreasing substantially the computation time. Also, for Algorithm~\ref{al:1}, once the procedure as been run once, it can be used to evaluate the coverage at any future data set without further calibration simulation (\textit{i.e.}, just the simulation at the new data).
%AE1 comment 1.
However although the parallelism in particular is very helpful, it is sometimes the case that the approximation we want to use cannot be made asymptotically exact, that is, we have no family of approximations with a ``resolution'' or ``sample size'' we can vary to improve accuracy but just a single ``fixed approximation''. The Ising model example in Section \ref{sec:part} illustrates this. We replace the partition function with a different function corresponding to the partition function for a different boundary condition. We have no practical way to improve this approximation. Where this is the case any serious analysis must provide some measure of the impact of the approximation on the reliability of results. A measure of the kind we provide, which measures the damage done to coverage, directly addresses the impact of the approximation on a quantity central to the analysis. In this setting a check of the sort we suggest may be worthwhile.
In all our examples, the likelihood only is approximated. In the notation above the approximate posterior may involve an approximation to the prior, the likelihood, or both.
\cite{talts18} give an example with an approximate prior. Cali\-brating a posterior based on an approximate prior $\tilde \pi(\theta)$ is a straightforward variant of our approach, so long as we can simulate the true prior $\phi\sim \pi(\cdot)$.

The remainder of the paper is structured as follows. We state our coverage estimation problem and describe two algorithms to solve it in Section \ref{sec:cal}. We show how they work on a very simple Gaussian model with a tempered likelihood in Section \ref{sec:ne}. We describe a methodology to correct credible intervals in Section \ref{sec:recal}, building on \cite{rodrigues18}. %See Section~\ref{sec:recal}, Equation~\ref{eq:Rrecal} and below for a discussion of the connection.
We give three further examples: an Ising model with a pseudo-likelihood in Section \ref{sec:part} illustrates all the methods in a simple setting where we have a sufficient statistic; a mixture model is analysed with a Variational Bayes approximation in Section \ref{sec:mix}; in Section~\ref{sec:car} we calibrate the coverage of a random partition in a Dirichlet-Process model for the distribution of random effects in a hierarchical model. The code generating results in Sections~\ref{sec:mix} and \ref{sec:mix} is available in is available in the online supplementary material \cite{LNRSupp19}.

\section{Estimating coverage under an approximation} \label{sec:cal}

Let $\tilde C_{y}$ and $C_{y}$ be level $\alpha$ credible sets for $\tilde\pi(\theta|y)$ and $\pi(\theta|y)$ respectively. These could for example be highest posterior density (HPD) sets (as in the examples in Sections~\ref{sec:ne} and \ref{sec:car}) or equal- or lower tailed intervals (as in the examples in Sections~\ref{sec:part} and \ref{sec:mix}).
If $\pi(\theta|y)=\tilde\pi(\theta|y)$ for all $\theta\in \Omega$ then $\tilde C_{y}=C_{y}$ is an exact credible set for $\phi$ when $\phi\sim\pi(\cdot)$ and $y\sim p(\cdot|\phi)$, that is
\[
\Pr(\phi\in C_{Y}|Y=y)=\alpha.
\]
In our approximation we take $\tilde C_{y}$ as an approximate level $\alpha$ credible set for $\pi(\phi|y)$. In this case we refer to
$\alpha$ as the nominal level. Denote by $b(y)$,
\begin{equation}\label{eq:b}
b(y)=\Pr(\phi\in \tilde C_{Y}|Y=y),
\end{equation}
this {\it operational} coverage probability.
%\marginpar{remark could define $b(y)$ in terms of $m(\phi,y',\theta)$}

We have additional Monte Carlo error if we use an estimate $\hat C_{y}(\theta)$ for $\tilde C_{y}$ computed using samples
$\theta=(\theta_{1},\ldots,\theta_{J})$ simulated so that $\theta_{j}\sim \tilde\pi(\cdot|y)$ for $j=1,\ldots,J$ (an abuse of notation as $\theta$ had $J=1$ up to this point).
%AE1 Typos section/comment 2
Let $c(y)$ give this second {\it realised} coverage probability at $Y=y$, averaged over $\phi$ and $\theta$, so that
\begin{equation}\label{eq:c}
c(y)=\Pr(\phi\in \hat C_Y(\theta)|Y=y).
\end{equation}
In this paper we give methods for estimating $b(y)$ and $c(y)$. In the examples in Sections~\ref{sec:ne}, \ref{sec:part} and \ref{sec:mix} we
compute or estimate $b(y)$, as $\tilde \pi(\theta|y)$ is relatively simple and we can compute $\tilde C_{Y}$. In the example in Section~\ref{sec:car} we
estimate $c(y)$.

We now give the estimators for $c(y)$. Estimators for $b(y)$ are similar but simpler as the estimate $\hat C_y$ is replaced by the $\tilde C_y$ (exact for $\tilde \pi(\theta|Y)$).
Let $Q(\phi)$ be a test distribution which we discuss below.
For $i=1,\ldots,M$ we simulate $\phi_{(i)}\sim Q(\cdot)$, $y_{(i)}\sim p(\cdot|\phi_{(i)})$ and $\theta_{(i)}=(\theta_{i,1},\ldots,\theta_{i,J})$ with $\theta_{i,j}\sim \tilde\pi(\cdot|y_{(i)})$ for $j=1,\ldots,J$. Here $y_{(i)}\in\mathcal Y$ is an entire data vector and similarly $\phi_{(i)}\in\Omega$ and $\theta_{(i)}\in \Omega^J$ for $i=1,\ldots,M$. We form an estimate $\hat C_{(i)}=\hat C_{y_{(i)}}(\theta_{(i)})$ of the approximate credible set using the sample set $\theta_{(i)}$ and use it to compute binary values
\[c_i=\mathbb{I}_{\phi_{(i)}\in \hat C_{(i)}}.\]
We have two natural  choices for estimating $c(y)$ from the ``data'' $(c_i,y_{(i)})_{i=1,\ldots,M}$ with different strengths and weaknesses.
Before we give these estimators we give an idealised, but impractical, algorithm estimating $c(y)$ consistently. See Algorithm~\ref{al:0}.
\begin{algorithm}
\caption{(in general unrealisable) estimation of realised coverage $c(y)$.}
\begin{algorithmic}[1]
  \FOR{ $i=1,\ldots,M$}
	\STATE Simulate $\phi_{(i)}\sim \pi(\cdot|y)$ and $\theta_{(i)}=(\theta_{i,1},\ldots,\theta_{i,J})$ with $\theta_{i,j}\sim \tilde\pi(\cdot|y)$ for $j=1,\ldots,J$.
	\STATE Estimate a credible set $\hat C_{(i)}=\hat C_{y}(\theta_{(i)})$ from the posterior samples, and binary values $c_i=\mathbb{I}_{\phi_{(i)}\in \hat C_{(i)}}$.
\ENDFOR	
  \STATE The estimated coverage is $\hat c(y)=M^{-1}\sum_i c_i$.
\end{algorithmic} \label{al:0}
\end{algorithm}
In this algorithm we simulate $\phi_{(i)}\sim\pi(\cdot|y)$ for $i=1,\ldots,M$, set $y_{(i)}=y$ and then simulate $\theta_{(i)}$, $\hat C_{(i)}$ and $c_i$ as above.
In this case our data is $(c_i,y)_{i=1,\ldots,M}$ and $\hat c=M^{-1}\sum_i c_i$ is unbiased and consistent for $c(y)$. Of course this is no use as we cannot simulate $\pi(\phi|y)$.
Algorithm~\ref{al:0} helps make clear what realised coverage means. We used this algorithm in the examples in Sections~\ref{sec:mix} and \ref{sec:car}
to demonstrate our estimators were working. %This is possible also but considerably more difficult for the example in Section~\ref{sec:part}.

We now give the estimators.
The first method we describe is logistic regression. The test distribution is $Q(\phi)=\pi(\phi)$ so the simulation step is that of \cite{cook06} and \cite{yao18}.
See Algorithm~\ref{al:1}.
In the triple $(\phi,y',\theta)$, we have $m(\phi,\theta|y')=\pi(\phi|y')\tilde\pi(\theta|y')$ conditionally, so if we take any particular $y'$ we
cover $\phi\in \hat C_{y'}(\theta)$ with probability $c(y')$. We take a vector $s(y)\in \R^p$ of $p$ summary statistics computed on the data and a vector $\gamma\in\R^p$ of regression parameters, and carry out logistic regression with
$\tilde c(y')=\mbox{logistic}(s(y')\cdot\gamma)$ and $c_i\sim {\rm Bernoulli}(\tilde c(y_{(i)}))$ independent observations for $i=1,\ldots,M$. Our coverage estimate is simply
$\hat c(y)=\mbox{logistic}(s(y)\cdot\hat\gamma)$ with $\hat\gamma$ the maximum likelihood estimator for $\gamma$. We found replacing linear logistic regression with a semi-parametric generalised additive model (a GAM) using methods outlined in \cite{wood11} worked well in our examples. The vector $s$ of summary statistics must be chosen with care. The examples in Sections~\ref{sec:ne} and \ref{sec:part} have a sufficient statistic so the choice of $s$ is straightforward, and more generally we expect good results for exponential family models. In Section~\ref{sec:mix}, the ABC-optimal rule given in \cite{fearnhead12} inspired the choice of $s$. Our regression approach did not give sensible estimates for a harder problem we tried (a large scale version of the example in Section \ref{sec:car}). For high dimensional data vectors $y\in \R^n$ with $n$ large the simulated data $y'$ do not enclose the real data $y$ and so we are making a large extrapolation of the coverage function $c(y)$.

\begin{algorithm}
\caption{Estimation of realised coverage $c(y)$ using logistic regression.}
\begin{algorithmic}[1]
  \FOR{ $i=1,\ldots,M$}
	\STATE Simulate $\phi_{(i)}\sim \pi(\cdot), y_{(i)}\sim p(\cdot|\phi_{(i)})$ and $\theta_{(i)}=(\theta_{i,1},\ldots,\theta_{i,J})$ with $\theta_{i,j}\sim \tilde\pi(\cdot|y_{(i)})$ for $j=1,\ldots,J$.
	\STATE Estimate a credible set $\hat C_{(i)}=\hat C_{y_{(i)}}(\theta_{(i)})$ from the posterior samples, and binary values $c_i=\mathbb{I}_{\phi_{(i)}\in \hat C_{(i)}}$.
\ENDFOR	
  \STATE Take $p$ summary statistics on the data $s:{\mathcal Y}\rightarrow \R^p$. Carry out logistic regression of $c_i\sim s(y_{(i)})$ onto the data yielding
  regression coefficient $\hat\gamma$.
  \STATE The estimated coverage is $\hat c(y)=\mbox{logistic}(s(y) \cdot \hat\gamma)$.
\end{algorithmic} \label{al:1}
\end{algorithm}

The second method we describe is importance sampling (IS) with proposal distribution $Q(\phi)=\tilde\pi(\phi|y)$. Denote by $\delta(y,y')$ a distance function in the space of data $\mathcal Y$. For small $\rho>0$ with $\Delta_y=\{y'; \delta(y,y')\le \rho\}$ we begin by making
an ABC-style approximation to $c(y)$. Define a probability function $d(y')$ for $\phi\sim \pi(\cdot)$ and $Y\sim p(\cdot|\phi)$ as
\begin{eqnarray}
d(y)&=& \Pr(\phi\in \hat C_{Y}(\theta)|Y\in \Delta_y) \label{eq:ist1}\\
 &=& \int_{\Omega\times \Omega^J}\int_{\mathcal Y} \mathbb{I}_{\phi\in \hat C_{y'}(\theta)} \frac{\pi(\phi)p(y'|\phi)\mathbb{I}_{y'\in \Delta_y}}{\Pr(Y\in \Delta_y)}\tilde\pi(\theta|y')d\theta dy' d\phi\label{eq:ist2} %\\
% &=& \int_{\Omega}\int_{\mathcal Y} \mathbb{I}_{\phi\in \hat C_{y'}(\theta)} p(y'|Y'\in \Delta_y)\pi(\phi|y')\tilde\pi(\theta|y)d\theta.
\end{eqnarray}
Equation~\ref{eq:ist2} uses the same abuse of notation we made in Equation~\ref{eq:c}, since $\theta$
is again a generic set of $J$ samples, equivalent to one of the sample sets $\theta_{(i)}, i=1,\ldots,M$ in Algorithm~\ref{al:2} below.
Also $\tilde\pi(\theta|y')$ represents the joint distribution of these $J$ samples. For example, if $\theta$ is the first $J$ samples output
by an MCMC run, then $\tilde\pi(\theta|y')$ gives their joint distribution in Equation~\ref{eq:ist2}.

Our plan is to estimate $d(y)$ in Equation~\ref{eq:ist2} using importance sampling, and then use this as an estimate for $c(y)$, the operational coverage of interest.
We motivate our approach by describing an approach that did not work in our setting. We might simulate
\begin{equation}\label{eq:pranglestyle}
(\phi,y',\theta)\sim \pi(\phi)p(y'|\phi)\mathbb{I}_{y'\in \Delta_y}\tilde\pi(\theta|y'),
\end{equation}
%by sampling
%\[(\phi_{(i)},y_{(i)})\sim \pi(\phi_{(i)})p(y_{(i)}|\phi_{(i)})\mathbb{I}_{y_{(i)}\in \Delta_y}\]
using rejection with $\phi\sim \pi(\cdot)$ and $y'\sim p(\cdot|\phi)$, and keeping only pairs $(\phi,y')$ satisfying $y'\in \Delta_y$, and then $\theta\sim \tilde\pi(\cdot|y')$ as before. This approach is used in the ABC-setting of \cite{prangle14} and characterises our different aims and methods. While \cite{prangle14} start with Equation~\ref{eq:iem} and then restrict to $y'\in \Delta_y$ in order to stop the prior-approximation (the $a=0$ case in Equation~\ref{eq:iep}) satisfying a coverage test,
we start with Equation~\ref{eq:iec} and aim to estimate the operational coverage. For the purpose of removing the $a=0$ solution to Equation~\ref{eq:iep} it may be enough to
take a rather large set $\Delta_y$. However, for estimating $c(y)$, we need simulated data close
to the real data. We would like the coverage $c(y')$ to be flat over $y'\in \Delta_y$,
so that in turn $d(y)\simeq c(y)$ is a reasonable approximation. For high dimensional simulated
data $y'$ we do not hit $\Delta_y$ in the rejection stage if we use Equation~\ref{eq:pranglestyle}.
We therefore use importance sampling $\phi\sim Q$ with proposal distribution $Q(\phi)=\tilde\pi(\phi|y)$.
This pushes our $\phi$ values into areas of parameter space where the realised $y'$ values are much closer to the data $y$.
We weight samples $(\phi,y',\theta)$ using the normalised weight function
\[
w(\phi,y',\theta) \propto \tilde p(y|\phi)^{-1},
\]
in order to get a consistent estimator for $d(y)$.
This gives coverage Algorithm~\ref{al:2}. This works well on simple problems, and even on the harder problems set out in Section~\ref{sec:car} and~\ref{sec:part}. However, the harmonic estimator
proved to be too unstable for the biggest problems we tried (again, a problem related to the example in Section~\ref{sec:car} but involving a much larger data set).
Developing a better estimator based perhaps on Sequential Monte Carlo is an obvious next step.

%\begin{figure}
%\fbox{\parbox{\textwidth}{Algorithm 2. Importance sampler estimating coverage
%\begin{enumerate}
%  \item For $i=1,\ldots,M$, (1) while $\delta(y_{(i)},y)>\rho$, simulate $\phi_{(i)}\sim \tilde \pi(\phi|y), y_{(i)}\sim p(y_{(i)}|\phi_{(i)})$ then (2) simulate $\theta_{(i)}=(\theta_{i,1},\ldots,\theta_{i,J})$ with $\theta_{i,j}\sim \tilde\pi(\cdot|y_{(i)})$ for $j=1,\ldots,J$.
%  \item For $i=1,\ldots,M$, estimate a credible interval $\hat C_{(i)}=\hat C_{y_{(i)}}(\theta_{(i)})$ from the posterior samples $\theta_{(i)}$, binary values $c_i=\mathbb{I}_{\phi_{(i)}\in \hat C_{(i)}}$ and normalised importance weights $w_i \propto \tilde p(\phi_{(i)}|y)^{-1}$.
%  \item The estimated coverage is $\hat c(y)=\sum_i w_i c_i$.
%\end{enumerate}
%}}\label{fig:a2}\caption{}
%\end{figure}

\begin{algorithm}
\caption{Importance sampler estimating the realised coverage $c(y)$.}
\begin{algorithmic}[1]
  \FOR{ $i=1,\ldots,M$}
	\STATE while $\delta(y_{(i)},y)>\rho$, simulate $\phi_{(i)}\sim \tilde \pi(\cdot|y), y_{(i)}\sim p(\cdot|\phi_{(i)})$ then
	\STATE simulate $\theta_{(i)}=(\theta_{i,1},\ldots,\theta_{i,J})$ with $\theta_{i,j}\sim \tilde\pi(\cdot|y_{(i)})$ for $j=1,\ldots,J$.
\ENDFOR
\FOR{$i=1,\ldots,M$}
  	\STATE estimate a credible set $\hat C_{(i)}=\hat C_{y_{(i)}}(\theta_{(i)})$ from the posterior samples $\theta_{(i)}$, binary values $c_i=\mathbb{I}_{\phi_{(i)}\in \hat C_{(i)}}$ and normalised importance weights $w_i \propto \tilde p(y|\phi_{(i)})^{-1}$.
\ENDFOR
  \STATE The estimated coverage is $\hat c(y)=\sum_i w_i c_i$.
\end{algorithmic} \label{al:2}
\end{algorithm}

One promising choice of distance function for scalar $\theta$ (\textit{i.e.} $\Omega=\R$) that seems well-adapted to our setting was suggested to us by \cite{bardenet18}.
The ``Kolmogorov-Smirnov distance function'' $\delta(y',y)= \left\lVert\hat G_y-\hat G_{y'}\right\rVert_{\infty}$ (hereafter, KS-distance) is based on the posterior CDF $G_y(\theta)$ of $\tilde\pi(\theta|y)$ at $y$. We are going to
sample $\theta\sim \tilde\pi(\cdot|y')$ anyway, and these samples may be used to form an empirical CDF $\hat G_y$. The downside of this is that we must
simulate $\theta\sim \tilde\pi(\cdot|y')$ for all the data-vectors $y'$ we simulate, not just the ones that satisfy $y'\in\Delta_y$, since we
need these samples to compute $\delta(y',y)$ itself.

\section{Coverage of a Normal mean}\label{sec:ne}

The diagnostic tools we have described cannot be ``fooled'' in quite the same way checks based on the exchangeability of $\phi$ and $\theta$ in Equation~\ref{eq:iem} can be.
This point and some other strengths and weaknesses are illustrated by the following very simple example.
%For example, if the summary statistic $s(y)$ used in Algorithm 1 is sufficient, if Algorithm 1 fits a logistic regression using a GAM with penalized regression splines, and $c(y)$ itself is smooth enough to be well represented by a spline, then we expect $\hat c(y)$ to approximate $c(y)$ very well, in the limit $M\rightarrow\infty$.
%Now suppose $\tilde p(y|\phi)=1$, ie the approximation is so bad that the posterior is just the prior. In this case $\phi$ and $\theta$ are exchangeable, so a test based on Equation~\ref{eq:iem} will not detect an error. However $c(y)$
%Since $c(y)$ really is the operating coverage probability at $y$, it Now, if the $\tilde\pi\ldots$
%converge to the exact coverage ($\hat c\rightarrow c(y)$ in the case of Algorithm 1, and $\hat c \rightarrow d(\Delta_y)$ for Algorithm 2).
%However they have their own weaknesses.

Suppose the prior is $\phi\sim \N(0,1)$, the observation model is $y\sim \N(\phi,1)$ (\textit{i.e.} the data vector is a scalar, we have just a single normal observation),
so that the exact posterior is $\pi(\phi|y)=\N(\phi;\frac y2, \frac12)$. Suppose now the approximate model has a tempered likelihood,
\begin{equation}
\tilde\pi(\theta|y)\propto \N(\theta; 0,1)\N(y;\theta,1)^{v},
\label{eq:nap0}\end{equation}
for some $v\geq 0$, that is,
\begin{equation}\tilde\pi(\theta|y)=\N\left(\theta; \frac{vy}{1+v}, \frac{1}{1+v}\right).\label{eq:nap}\end{equation}
The approximation is good when $v=1$ (no approximation) and bad when $v=0$ (the approximation $\tilde\pi(\theta|y)$ coincides with the prior $\pi(\theta)$).

\begin{figure}
  % Requires \usepackage{graphicx}
  \centering
  \hspace*{-0.25in}\includegraphics[width=6in]{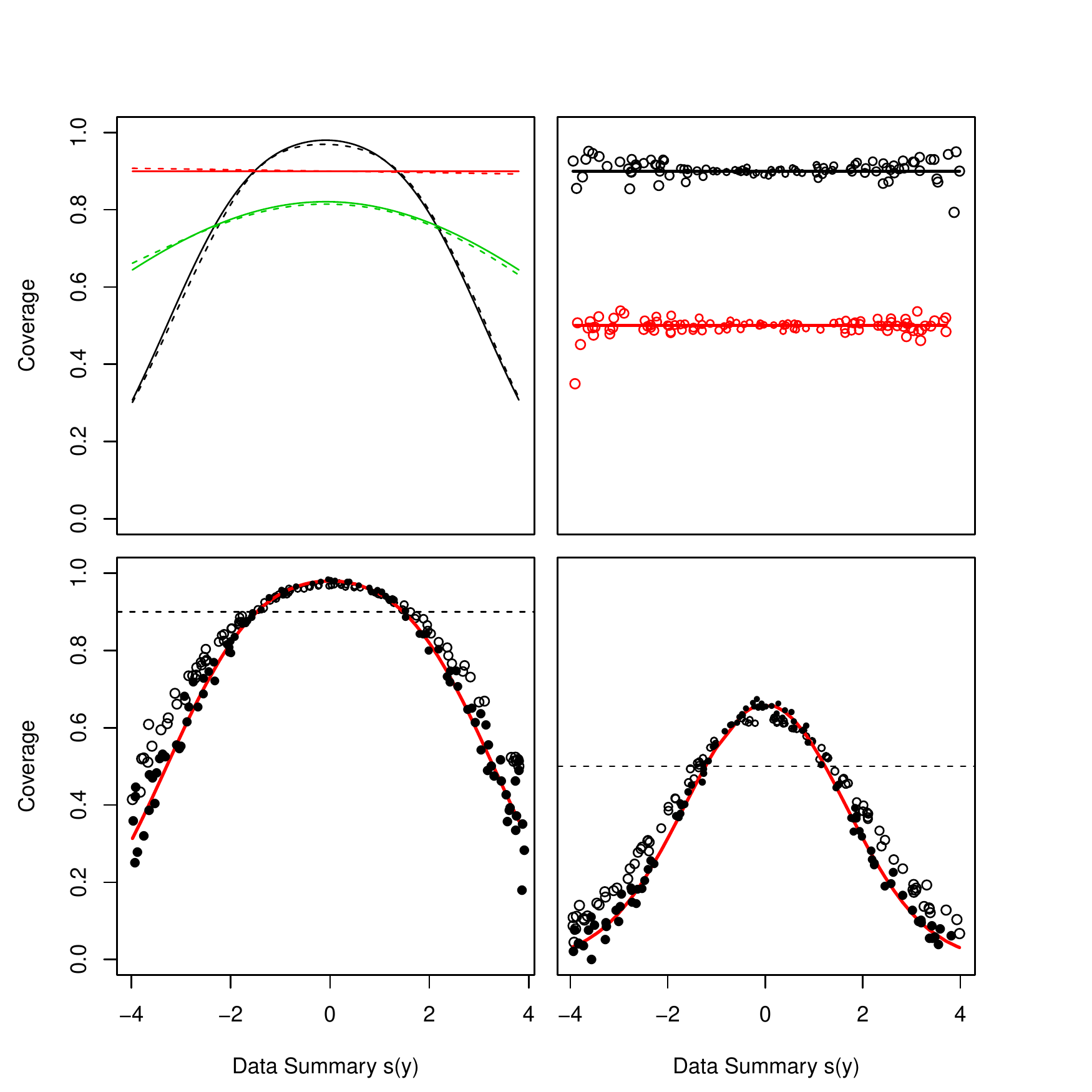}
  \caption{Tempered normal approximate likelihood. $Y$-axis is operational coverage. $X$-axis is scalar summary statistic $s(y)=y$. (Top Left) logistic regression (GAM) estimate of operational coverage $b(y)$ when $\alpha=0.9$. Line types: truth $b(y)$, solid; estimated $\hat b(y)$, dashed.
Colors: Red, $v=1$, Green $v=0.5$, Black $v=0$. (Top Right) IS estimates of $b(y)$ at $v=1$: lines give exact $b(y)$ at $\alpha=0.9$ (black) and $\alpha=0.5$ (red); points give IS estimates of $b(y)$, larger points have lower ESS. (Bottom Left) IS estimates of operational coverage with $v=0$ and $\alpha=0.9$: red line gives true operational coverage; open circles $\rho=1$, full circles $\rho=0.3$. (Bottom Right) As bottom left but with $\alpha=0.5$.}\label{fig:ne}
\end{figure}

%Let $\tilde C_{y}$ be a level $\alpha$ credible set for $\tilde\pi(\theta|y)$
%and let
%\[
%b(y)=\Pr(\phi\in \tilde C_Y|Y=y)
%\]
%so that $b(y)$ is the exact coverage achieved by our approximate posterior $\tilde\pi(\theta|y)$, without Monte Carlo error.
Let $Z_\alpha$ and $\Phi(z)$ be respectively the $\frac{1-\alpha}{2}$ quantile (recall, in this paper $\alpha$ is typically $0.9$ or $0.95$) and CDF of a standard normal and let
\[
B_\pm=\frac{vy}{1+v}\pm Z_{\alpha}\sqrt{\frac{1}{1+v}}
\]
so that $\tilde C_y=[B_-,B_+]$ is a credible set for $\theta$ given $y$, with nominal level $\alpha$. Now, referring to the coverage probability $b(y)$
defined in Equation~\ref{eq:b}, we have
\[
b(y)=\Phi\left(\sqrt{2}(B_+-\frac y2)\right)-\Phi\left(\sqrt{2}(B_--\frac y2)\right),
\]
so that $b(y)=\alpha$ when $v=1$ and
%\[
%b(y)=\Phi(\sqrt{2}(Z_{\alpha/2}-y/2))-\Phi(-\sqrt{2}(Z_{\alpha/2}+y/2))
%\]
%when the approximate posterior is just the prior. This is
does not in general equal $\alpha$ otherwise (see Figure~\ref{fig:ne}).

Consider estimating $b(y)$ using Algorithms~\ref{al:1} and~\ref{al:2}.
The algorithms simplify slightly as we do not need to draw samples $\theta\sim \tilde\pi(\cdot|y')$ in order to form an
estimate $\hat C_{y'}$ for $\tilde C_{y'}$ at simulated date $y'$ and so we estimate $\hat b(y)$ in Equation~\ref{eq:b} rather than $\hat c(y)$ in Equation~\ref{eq:c}. In Figure~\ref{fig:ne}
we plot the exact operational coverage $b(y)$ and its estimate $\hat b(y)$ as functions of the summary statistic $s(y)=y$ for various values of $v$
and $\rho$ taking $M=10000$ simulated ``data points'' $(c_i,y_{(i)})_{i=1}^M$. At top left we see logistic regression with a GAM accurately estimates the operational coverage at all data and all approximation values.
The remaining graphs show the behavior of the IS-estimator, Algorithm~\ref{al:2}. Again we have $M=10000$ but now the data points are simulated using the importance sampling proposal distribution - the approximate posterior itself.
At top right the ``easy'' case $\alpha=0.9$ and $v=1$ (\textit{i.e.} no approximation) is in fact demanding as the estimator in Algorithm~\ref{al:2} is now unbiased but sharply skewed at large $\alpha$.
%AE1 comment 2 Oh yes that's true. Biased for b -> not consistent for b (as consistent for d and d ne b in general)
At bottom left and right we see evidence that IS is not consistent for $b(y)$ as it is a consistent for $\Pr(\phi\in \tilde C_Y|Y\in \Delta_y)$ not $\Pr(\phi\in \tilde C_Y|Y=y)$.
However, the bias is reduced as $\rho$ is decreased.
The two methods, logistic regression and IS, expose the poor approximation at $v=0$ (approximation-is-prior, corresponding to $a=0$ in Equation~\ref{eq:iep}) very well.
Notice that a test based on measuring exchangeability of $\phi$ and $\theta$ would not show up this poor approximation.

%AE1 comment 3 - OK I now claim we show the variance is finite not just the variance of the delta-method estimate!
In Appendix A we show that if $0\le v < 2$ then, for this simple normal example,
the un-normalised IS estimator
in Algorithm~\ref{al:2} satisfies a Central Limit Theorem.
If the approximate posterior $\tilde\pi(\theta|y)$ is under-dispersed with respect to the true posterior $\pi(\theta|y)$
then the performance of the simple IS estimator we are using may be poor. We expect this property to hold in a qualitative sense in
other settings.

\section{Achieving the nominal level}\label{sec:recal}

The material in this section can be omitted at first reading. It is of independent interest, and highlights the connection between our work and \cite{rodrigues18}.
We have estimated the operational coverage $b(y)$ and the realised coverage $c(y)$ at the data for general credible sets respectively $\tilde C_{y}$ and $\hat C_y$ of fixed nominal level $\alpha$.
%\[
%c(y)=\Pr(\phi\in \hat C_{y'}(\theta)|y'=y)
%\]
%with $\phi\sim \pi(\phi)$, $y'\sim p(y'|\phi)$ and $\theta\sim \pi(\theta)\tilde p(y'|\theta)$.
%giving a regression based estimate for $c(y)$ itself, and a consistent importance sampling estimator for $\Pr(\phi\in \hat C_{y'}(\theta)|y'\in \Delta_y)$.
The framework above does not require the parameter $\theta$ to be a real scalar.
We now restrict to $\Omega=\R$ and consider the level-$\alpha$ dependence of $c(y)$ specifically for lower tail credible intervals.
%We will return to the more general setting later in this section.

Let $\tilde q_{y'}(\alpha)=G^{-1}_{y'}(\alpha)$ be the level $\alpha$ quantile of $\tilde \pi(\theta|y')$ where $G_{y'}(\theta)=\int_{-\infty}^\theta \tilde \pi(\theta'|y')\, d\theta'$ is the CDF of $\theta$ at $y'$ in the approximate posterior. Let $F_{y'}(\phi)=\int_{-\infty}^\phi \pi(\phi'|y')\, d\phi'$ be the CDF for the true posterior given generic data $y'$. The operational coverage we achieve with our approximation $\tilde\pi(\theta|y)$ is a function of $\alpha$ at each $y$-value, and we write this as
\begin{equation}\label{eq:by}
b_y(\alpha)=\Pr(\phi\le \tilde q_{Y}(\alpha)|Y=y).
\end{equation}
This is the same as $b(y)$ but the dependence on the nominal level of coverage $\alpha$ is explicit.
Equation~\ref{eq:by} is the relation $b_y(\alpha)=F_y\circ G^{-1}_y(\alpha)$. This can be inverted to give the map from $G_{y}$ to $F_{y}$ at the data $y$,
\[
F_y(\phi)=b_y\circ G_{y}(\phi).
\]
Our coverage function $b_y(\alpha)$ is just the ``distortion function'' mapping the approximate CDF to the true CDF at the data.
%There is a corresponding realised coverage function $c_y(\alpha)$ in the case where we estimate $\hat q_{y'}(\alpha)=\theta_{\lceil (1-\alpha) J\rceil}}$
%using the $\lceil (1-\alpha) J\rceil$-order statistic of $J$ samples drawn from $\tilde\pi(\theta|y')$, that is
%\begin{equation}\label{eq:cy}
%c_y(\alpha)=\Pr(\phi\le \hat q_{Y}(\theta;\alpha)|Y=y).
%\end{equation}
If we form estimates $\hat b_y(\alpha)$ of $b_y(\alpha)$ (using Algorithm~\ref{al:1} or~\ref{al:2}) and $\hat{G_{y}}$ (the empirical CDF obtained using for example MCMC targeting $\tilde \pi(\theta|y)$) we may ``recalibrate'' $G_{y}$ at the data $y$ to better estimate $F_y$ using the estimator
\begin{equation} \label{eq:recal}
\hat F_y(\phi) = \hat b_y \circ \hat{G_{y}}(\phi).
\end{equation}
%\marginpar{oh bloody hell this should be $1-G$ I think as $b$ acts on $\alpha$ not $1-\alpha$}
This idea is set out in \cite{rodrigues18} who use it to map $\phi|y'$ to $\phi|y$ via the adjustment $\phi^{(adj)}=G_y^{-1}\circ G_{y'}(\phi)$. In our setting this sort of map
would be effective. When we approximate $b_y(\alpha)$ with $\Pr(\phi\le \tilde q_{Y}(\alpha)|Y\in \Delta_y)$ we assume $b_{y'}(\alpha)$ does not depend on $y'$ for
$y'\in \Delta_y$. If $\phi\sim \pi(\cdot)$ and $y'\sim p(\cdot|\phi)$ so that $\phi|y'\sim \pi(\cdot|y')$ then
\begin{eqnarray*}
% \nonumber to remove numbering (before each equation)
  \phi^{(adj)} &=& G_y^{-1}\circ G_{y'}(\phi)\nonumber \\
               &=& F_y^{-1}\circ b_y\circ b^{-1}_{y'}\circ F_{y'}(\phi)\nonumber \\
          &\simeq& F_y^{-1}\circ F_{y'}(\phi),\label{eq:Rrecal}
\end{eqnarray*}
as $b_y\circ b^{-1}_{y'}(x)=x$ if $b_{y'}$ does not depend on $y'$.
After adjustment \cite{rodrigues18} have $\phi^{(adj)}\sim \pi(\cdot|y)$ (approximately). It is straightforward to check that $b_{y}$ is invertible at each $y$. \cite{rodrigues18} use the empirical estimate $\hat G_y^{-1}\circ \hat G_{y'}(\phi)$ to implement the map.

%This approach does not work in an importance sampling setting
%as we will draw $\tilde\phi\sim \tilde\pi(\tilde \phi|y)$ and $y'\sim p(y'|\tilde \phi)$ so we have draws
%$\tilde\phi|y'\sim \pi(\tilde \phi|y')\tilde p(y|\tilde \phi)$ rather than $\phi\sim \pi(\phi|y')$, and a different approach to calibration.
%Note that [Rodriguez et al 17] work in an ABC setting so
%typically $y$ is replaced by summary statistics $s(y)$ and they map $\phi|s(y')$ to $\phi|s(y)$ etc. Also, the recalibration map in [Rodriguez et al 17]
%is, like our calibration map in Equation~\ref{eq:recal}, based on empirical CDF estimates $\hat G_y$ etc.

We do not wish to make an adjustment of the kind \cite{rodrigues18} make, as we do not need to map samples $\theta$ at $y'$ to samples $\phi^{(adj)}$ at the data $y$. We are interested in cases where we
can generate approximately distributed samples at $y$ by sampling $\theta\sim\tilde \pi(\cdot|y)$ itself. These samples could be recalibrated (at $y$)
using Equation~\ref{eq:recal}.
For example, if we seek a corrected median estimate we can replace the median estimate $\hat G_y^{-1}(0.5)$ for $\tilde\pi(\theta|y)$
with $\hat G_y^{-1}\circ\hat b_y^{-1}(0.5)$.
Our aim is to provide an estimate $\hat c$ of the coverage of an approximate credible set $\hat C_y$, not an improved credible set.
However we show how the correction may be made and give an example in Section~\ref{sec:part}.

Given Monte Carlo samples $\theta=(\theta_{1},\ldots,\theta_{J})$ distributed as $\tilde\pi(\cdot|y')$ (notation as Equation~\ref{eq:ist2}),
and ordered so that $\theta_{i,j}<\theta_{i,j+1}$ for $j=1,...,J-1$, we estimate $\hat q_{y'}(\theta;\alpha)=\theta_{(\lceil \alpha J\rceil)}$. The realised coverage is
\[
c_{y,J}(\alpha)=\Pr(\phi\le \hat q_{Y}(\theta;\alpha)|Y=y).
\]
%This time we have
%\[
%\hat q_{y'}(\alpha,\theta)=\hat F^{-1}_{y}(\alpha;\theta),
%\]
%where
%\[
%\hat F_{y}(\phi;\theta) = J^{-1}\sum_{j=1}^J \mathbb{I}_{\phi\le \theta_j},
%\]
%and
%\[
%\hat F^{-1}_{y}(\alpha;\theta) = \min_{\phi\in \Omega}\{\phi: \hat F_y(\phi;\theta)\ge \alpha\},
%\]
%so we have
%\[
%c_y(\alpha)=E(F_y\circ \hat F^{-1}_{y}(\alpha;\theta)),
%\]
%where the expectation is taken over $\theta$.
Again we assume $c_{y,J}(\alpha)\simeq d_{y,J}(\alpha)$
where now
\[
d_{y,J}(\alpha)=\Pr(\phi\le \hat q_{Y}(\theta;\alpha)|Y\in\Delta_y)
\]
is a function of the level.
%Remark: let $r(\phi,\theta)=\sum_{j=1}^J \mathbb{I}_{\phi \le \theta_j}$. Then $r(\phi,\theta)\le \lceil \alpha J\rceil$ if and only if $\phi\le \hat q_y(\alpha,\theta)$ so this is just another way of writing the same thing.
Reasoning as before,
\[
d_{y,J}(\alpha)=\int_{\Omega\times \Omega^J}\int_{\mathcal Y} \mathbb{I}_{\phi\le \hat q_{y'}(\theta;\alpha)} \frac{\pi(\phi)p(y'|\phi)\mathbb{I}_{y'\in \Delta_y}}{\Pr(Y'\in \Delta_y)}\tilde\pi(\theta|y')d\theta dy' d\phi.
\]
We estimate this in Algorithm~\ref{al:3} using importance sampling draws from $\phi_{(i)}\sim\tilde\pi(\cdot|y_{(i)})$ and weighting by $1/\tilde p(y|\phi_{(i)})$ as before.
We are estimating $c_{y,J}(\alpha)$ via a consistent estimator for $d_{y,J}(\alpha)$.
In this setting it seems clear that following \cite{bardenet18} and using the distance function $\delta(y',y)= \left\lVert\hat G_y-\hat G_{y'}\right\rVert_{\infty}$ is desirable:
if the CDF's are similar, at least for $y'\in\Delta_y$, then we may hope that the distortion functions $c_{y,J}(\alpha)$ and $c_{y',J}(\alpha)$ are similar,
supporting our assumption $c_{y,J}(\alpha)\simeq d_{y,J}(\alpha)$.
\begin{algorithm}
\caption{Importance sampler estimating the realised coverage function $c_y(\alpha)$.}
\begin{algorithmic}[1]
  \FOR{ $i=1,\ldots,M$}
	\STATE while $\delta(y_{(i)},y)>\rho$, simulate $\phi_{(i)}\sim \tilde \pi(\cdot|y), y_{(i)}\sim p(\cdot|\phi_{(i)})$ then
	\STATE simulate $\theta_{(i)}=(\theta_{i,1},\ldots,\theta_{i,J})$ with $\theta_{i,j}\sim \tilde\pi(\cdot|y_{(i)})$ for $j=1,\ldots,J$, ordered so that $\theta_{i,j}<\theta_{i,j+1}$ for $j=1,\ldots,J-1$.
\ENDFOR
\FOR{$i=1,\ldots,M$}
  	\STATE compute the step functions $c_i(\alpha)=\mathbb{I}_{\phi_{(i)}\le \theta_{i,\lceil \alpha J\rceil}}$ and normalised importance weights $w_i \propto \tilde p(y|\phi_{(i)})^{-1}$.
\ENDFOR
  \STATE The estimated coverage function at level $\alpha$ is $\hat c_y(\alpha)=\sum_i w_i c_i(\alpha)$.
\end{algorithmic} \label{al:3}
\end{algorithm}

Algorithm~\ref{al:3} makes Algorithm~\ref{al:2} redundant.
We can use the function $\hat c_y(\alpha)$ output by Algorithm~\ref{al:3} to estimate the realised coverage $c(y)$ of our estimate $\hat C_y(\theta)$
by evaluating $\hat c_y(\alpha)$ at the value of $\alpha$ used to form $\hat C_y(\theta)$.
However we can also correct $\hat C_y(\theta)$ to get a new interval with the required operational coverage. % \marginpar{what better properties?}}
If we find the value $\tilde\alpha$ say, satisfying $\alpha=c_y(\tilde\alpha)$ then $(-\infty, \tilde q_{y}(\tilde\alpha)]$ covers $\phi$
with probability $\alpha$. In practice we work with estimates, so we solve $\alpha=\hat c_y(\hat\alpha)$ and estimate the credible set
$\hat C_y(\theta)=(-\infty, \hat q_{y}(\theta;\hat\alpha)]$ based on an adjusted level, in order to make the realised coverage match the desired nominal coverage.
We give an example of this calculation in the next section (see the last paragraph and Figure~\ref{fig:ifr} of Section~\ref{sec:part}).
However, as noted above, this is a by-product of the analysis, not the essential point. We seek a quality guarantee, not a correction.

Algorithm~\ref{al:3}, given here for lower-tail intervals, can be extended to handle
equal-tailed intervals and HPD regions (\cite{rodrigues18} set this out in a general way). It is sufficient that
credible sets at smaller nominal $\alpha$ are nested within credible sets at larger $\alpha$.
%so $\alpha\le\alpha'$ implies $\hat C_y(\theta; \alpha) \subseteq \hat C_y(\theta; \alpha')$
That makes $c_i(\alpha)$ an increasing step function and $\hat c_y(\alpha)$ non-decreasing.

\section{Coverage of the Ising model smoothing parameter} \label{sec:part}

\begin{figure}
  \centering
\includegraphics[width=3in]{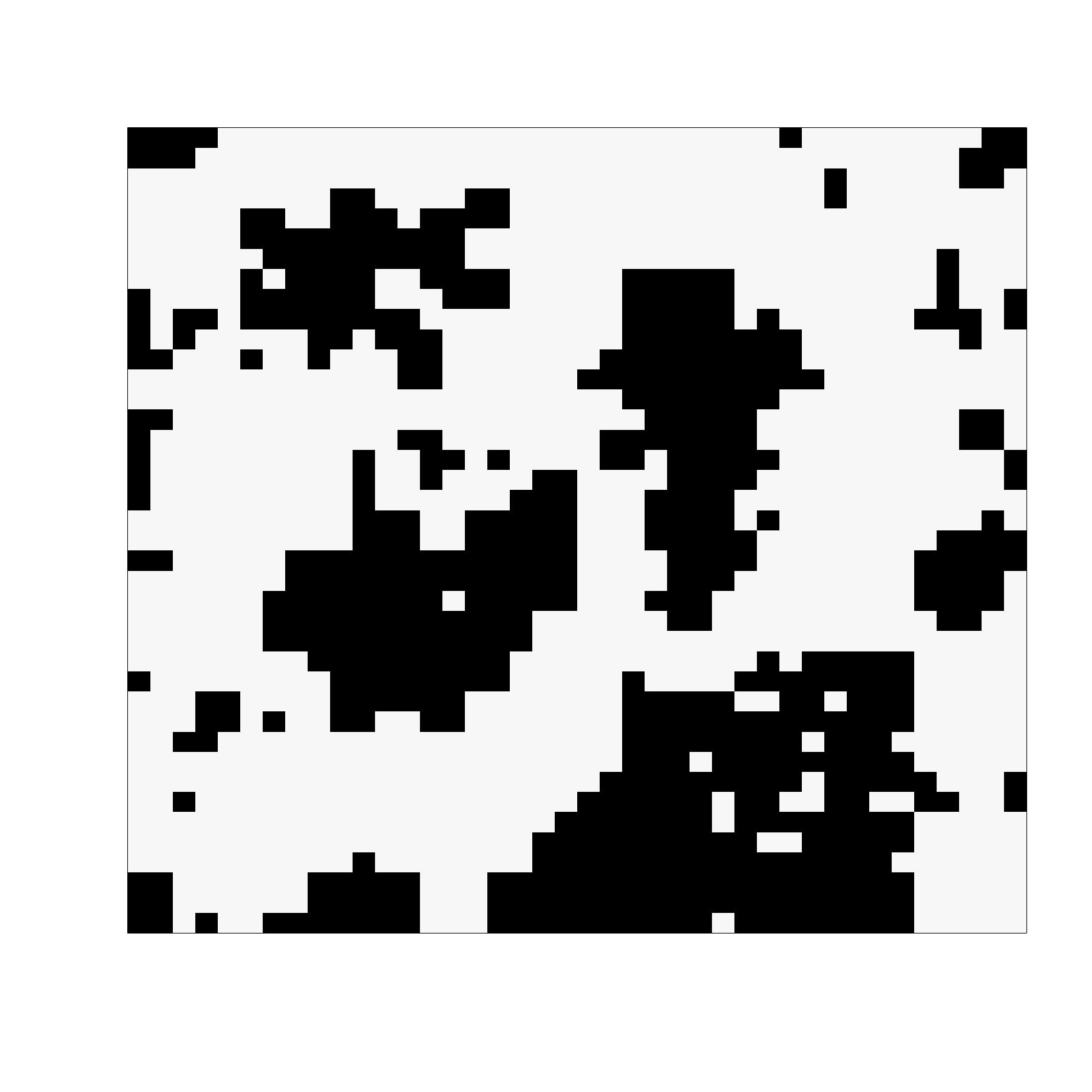}
\caption{Ice floe data from \protect\cite{bornn13}.}\label{fig:if}
\end{figure}

\begin{figure}
  \centering
  % Requires \usepackage{graphicx}
  %\[
  %\begin{array}{ccc}
  \includegraphics[width=3in]{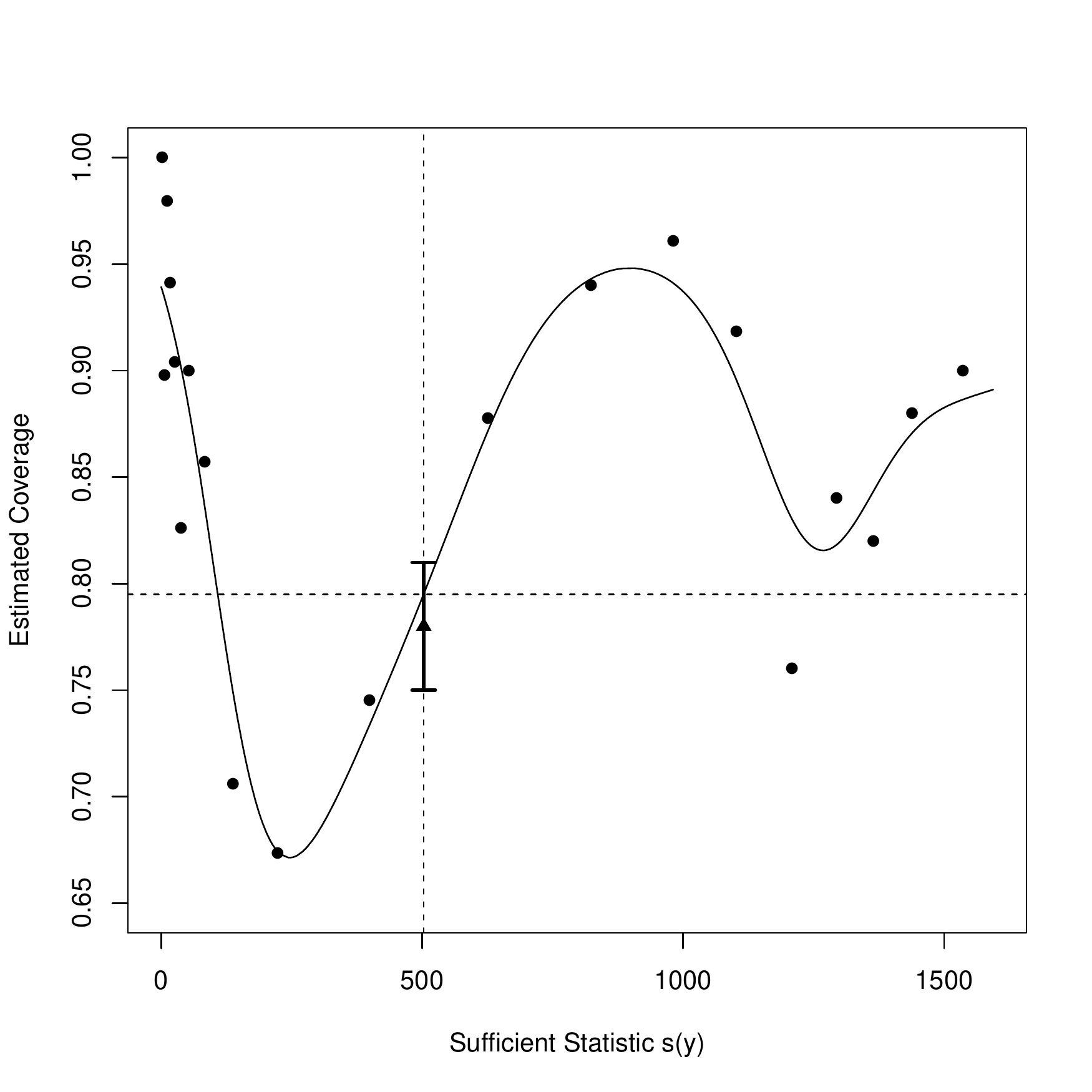}\includegraphics[width=3in, height=2.85in]{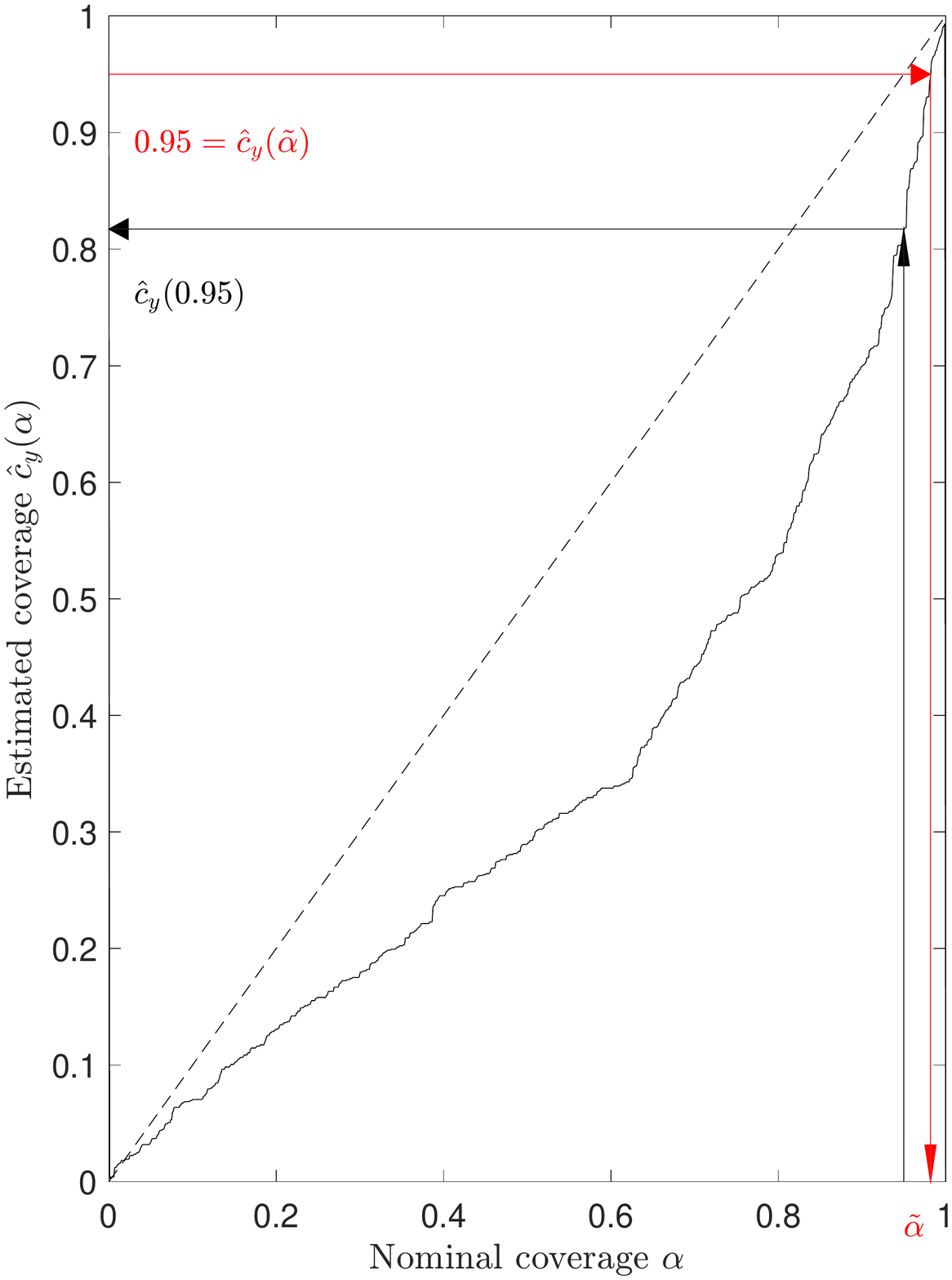}
  %\end{array}
  %\]
  \caption{Estimation of coverage for the Ising model of Section \ref{sec:part}. (Left) Estimated coverage as a function of the sufficient statistic $s(y)$: points are average coverage in $x$-axis bins each containing 50 data points; solid curve is GAM regression of coverage response on $s(y)$; vertical (horizontal) dashed line at the data value $s(y)=503$ (resp. estimated operational coverage $\hat c=0.80$)
  from the image at left; the error bar gives the IS-estimate $\hat c=0.78$ and error $\hat\sigma=0.03$. (Right) Solid line is estimated realised coverage $\hat c_y(\alpha)$ at data $y$ plotted against nominal or target coverage $\alpha$; dashed line is ideal realised coverage; black arrows give map from nominal coverage 0.95 to realised coverage 0.82; red arrows give inverse map from target realised coverage 0.95 to the nominal coverage $\tilde\alpha=0.98$ which would achieve it.}\label{fig:ifr}
\end{figure}

The image in Figure~\ref{fig:if} is a data set quoted from \cite{bornn13} where it was used to illustrate adaptive Wang-Landau simulation of
a binary Markov Random Field. Those authors registered it by thresholding a larger grey-level image of ice floes published in \cite{raftery92}.
We will fit a binary Markov random field (MRF) to these data, and illustrate our methods on the problem of estimating the smoothing parameter, $\phi$ (also
referred to as the inverse temperature, and usually denoted $\beta$, as we fit the Ising model). Here the data vector $y$ records a $N\times N$ square
array of binary values given by the grey level in the image in Figure~\ref{fig:if}, where $N=40$. In the true observation model the MRF observation model will have free boundary conditions. This is
a natural modelling choice but gives an intractable likelihood for $\phi$. We will approximate this with an observation model which has
periodic boundary conditions but is otherwise identical. The likelihood for this second model is easily evaluated to machine precision.
The MatLab code generating all results in this section is available in the online supplementary material \cite{LNRSupp19}.
Foreshadowing our results, Figure~\ref{fig:ifr} (left) shows the estimated coverage function $b(y)$, the probability our ``wrong'' credible set for $\phi$
with nominal coverage $\alpha=0.95$ covers $\phi$ if $\phi$ is a draw from the prior and the data-image $y$ is a draw from the observation model. In this problem the coverage depends
only on a scalar sufficient statistic $s(y)$ defined below, so in Figure~\ref{fig:ifr} (left) we have plotted $b(y)$ against
$s(y)$. We can see that the coverage of our estimated credible interval $\tilde C_{y'}$ varies significantly over the space $\mathcal Y$ of data sets $y'$.
However our estimation methods give good estimates of the operational coverage we are achieving at the value of $s(y)$ corresponding to the data in Figure~\ref{fig:if}.
\cite{zhu18} calibrate an approximate fit to a Potts model using a coarsening procedure related to the real-space renormalisation group methods \cite{Gidas89} applies in image processing. \cite{zhu18} compute the frequentist coverage probabilities at chosen values of the parameter $\phi$.

The Ising model is a well known Markov model for a binary random field. Let $G$ be a graph with edges $E$ and vertices $V$.
For $v\in V$ let $y_v\in \{0,1\}$ be binary data at vertex $v$. Let $y=(y_v)_{v\in V}$ so that $y\in \mathcal{Y}$ with $\mathcal{Y}=\{0,1\}^{N^2}$. Let $\bra u,v\ket\in E$ denote a generic edge
in $G$ with vertices $u,v\in V$. Denote by
\[f(y;E)=\sum_{\bra u,v\ket\in E} \mathbb{I}_{y_u\ne y_v}\]
the number of edges connecting non-equal neighbours on the graph. In our case $G$ is
a rectangular $N\times N$ lattice with $N=40$ and a free boundary, $G_F=(E_F,V)$ say. On this graph interior vertices have degree 4, edge vertices
have degree 3, and corner vertices have degree 2. We will consider also lattices $G_P=(E_P,V)$ with periodic or toroidal boundary conditions.
In this case the lattice is wrapped onto a torus and all vertices have 4 neighbours.

Let $\phi\ge 0$ be a positive scalar smoothing parameter. The Ising model distribution for a rectangular lattice with a free boundary is
\begin{equation}\label{eq:isingF}
p_F(y|\phi)=\frac{1}{Z_F(\phi)}\exp\left(-\phi f(y;E_F)\right)
\end{equation}
where
\begin{equation}\label{eq:isingFZ}
Z_F(\phi)=\sum_{x\in \mathcal{Y}} \exp\left(-\phi f(x;E_F)\right)
\end{equation}
is a normalising constant. The normalising constant $Z_F(\phi)$ is an intractable function of $\phi$, for free boundary conditions, for $N$ at all large.
However, it is available from \cite{beale96} in a simple closed form derived by \cite{Kaufman49} for the special case of periodic boundary conditions
(for MatLab implementation see \cite{LNRSupp19}).
%A MatLab function evaluating $\log(Z_P)$ is given in an appendix.

%$x$ be an $M\times N$ array, $x\in \{0,1\}^{MN}$ and let $E$ be a nWe use it to illustrate our calibrations schemes as it is
%simple and familiar but leads to an intractable likelihood for the Ising model smoothing parameter.

Consider the problem of estimating the smoothing parameter $\phi$ for the data in Figure~\ref{fig:if}. Values of $\phi$ greater than about 2
are uninteresting for image modeling purposes as the image is essentially all 0's or all 1's under the prior. We take as prior $\pi(\phi)\propto\mathbb{I}_{\phi\in [0,2]}$.
The posterior is
\[
\pi(\phi|y)\propto \frac{1}{Z_F(\phi)}\exp\left(-\phi f(y;E_F)\right)\mathbb{I}_{\phi\in [0,2]}.
\]
The likelihood for $\phi$ depends on the data $y$ through the scalar quantity $s(y)=f(y;E_F)$ only, so this statistic is sufficient.
This posterior is doubly intractable, due to the $Z_F(\phi)$-dependence. One approximate solution is to simply replace
$Z_F(\phi)$ with $Z_P(\phi)$, which we can compute.
%We expect this to be good for large lattices since
%\[
%f(y;E_P)=f(y;E_F)+\sum_{\bra u,v\ket\in E_P\setminus E_F} \mathbb{I}_{y_u\ne y_v}
%\]
%and there are $O(N^2)$ terms in the first term and $O(2N)$ in the second.
Denote by
\[
\tilde\pi(\theta|y)\propto \frac{1}{Z_P(\theta)}\exp\left(-\theta f(y;E_F)\right)\mathbb{I}_{\theta\in [0,2]}.
\]
the approximate posterior obtained on making this substitution. In this case the approximate
posterior density and its CDF are readily evaluated using the formulae for the partition function derived by \cite{Kaufman49}, and we can compute $\tilde C_{y}$ to machine precision.
%AE1 comment 5 ``readily evaluated''
The result for the data in Figure~\ref{fig:if} is $\tilde C_y=[0.84,0.90]$. This is exact for $\tilde\pi(\theta|y)$
but only approximate for $\pi(\phi|y)$.
%This is our approximate level $\alpha$ equal-tailed
%credible set for $\phi$ with $\alpha=0.95$.
%It satisfies
%\[
%\alpha=\int_0^2 \mathbb{I}_{\theta\in \tilde C_y} \tilde\pi(\theta|y)d\theta
%\]
We would like to know the operational coverage $b(y)$ this approximation achieves.

We now run Algorithms~\ref{al:1} and~\ref{al:2} to estimate $b(y)$.
As the credible interval under the approximation is available without simulation,
the algorithms again simplify (\textit{i.e.} as in Section~\ref{sec:ne} we estimate $b(y)$ rather than $c(y)$ in Equations~\ref{eq:b} and \ref{eq:c}).
%We do not need to simulate $\theta_{(i)}=(\theta_{i,1},\ldots,\theta_{i,J})$ for $i=1,\ldots,M$
%and estimate a credible interval $\hat C^{(y_{(i)})}(\theta_{(i)})$ from the posterior samples. We simply compute the credible interval
%$\tilde C_{(i)}=\tilde C^{(y_{{i}})}$ analytically and set $c_i=\mathbb{I}_{\phi_{(i)}\in \tilde C_{(i)}}$.
We made $M=1000$ simulations of $\phi$ and $y'\sim \tilde\pi(\cdot|\phi)$ in Algorithms~\ref{al:1} and~\ref{al:2}. For our algorithm to be correct this
simulation should be exact. However we simulated $y_{(i)}$ using a simple single-site MCMC algorithm with a very large run-length.
Although it is possible that this introduces another layer of bias, we took --- for the purpose of this analysis --- a very large
run length and checked convergence carefully so that bias involved is negligible compared to the effect due to the boundary condition.

In our logistic regression in Algorithm~\ref{al:1} we use, as a covariate in the logistic regression,
the summary statistic $s(y_{(i)})=f(y_{(i)};E_F)$ where $y_{(i)}\sim p(\cdot|\phi_{(i)})$.
%We tested, as a second covariate, $d^{(i)}=\max_\theta(|F_{y}(\theta)-F_{y^{(i)}}(\theta)|)$, the KS-statistic measuring the difference between the two
%approximate posterior distributions $\tilde\pi(\theta|y)$ and $\tilde\pi(\theta|y^{(i)})$. Notice that $d^{(i)}$
%is a deterministic function of $f(x^{(i)})$ - it is a function of $f(x^{(i)};E_P)$, and in any case the two regressors are not linearly related.
%In our linear regression this covariate was not significant (in the presence of the sufficient statistic) so we dropped it in forming our estimate
%for the regressed coverage.
In Figure~\ref{fig:ifr} (left) we plot the estimated operational coverage $b(y')$ as a function of the sufficient statistic $s(y')=f(y';E_F)$.
This is the coverage we get over data space if we aim at a fixed nominal coverage equal 0.95 (\textit{i.e.} $\alpha=0.95$ is fixed, as in Section~\ref{sec:cal}).
The curve is a semi-parametric logistic regression (a GAM computed using the R function $\tt gam()$ in the package {\tt mgcv}, see \cite{wood11})
of the coverage response $c_i$ in Algorithm~\ref{al:1} on the sufficient statistic,
where $y_{(i)}\sim p(\cdot|\phi_{(i)})$ is the simulated data at $\phi_{(i)}$ and $\phi_{(i)}\sim \pi(\cdot)$ is a draw from
the prior, for $i=1,\ldots,M$. In this setting, with a sufficient statistic, this is a fairly reliable estimate of the true operational coverage function $b(y)$,
interpolating the proportion of $c$-values equal to 1 in the neighbourhood of each $s(y)$-value.
The value of the sufficient statistic at the data is $s(y)=503$, so our best estimate of $b(y)$ at the real data $y$ (\textit{i.e.} the GAM fit at $s(y)$) is $0.80$.
As an aside, we note that if we estimate the operational coverage using linear logistic regression instead of a GAM in Algorithm~\ref{al:1}, we get 0.85,
though the nonlinear dependence in Figure~\ref{fig:ifr} suggests a local linear regression windowed on data close to $s(y)$ would be more reliable.

In Algorithm~\ref{al:2}, we used the KS-distance $\delta(y,y')=\left\lVert G_{y}-G_{y'}\right\rVert_{\infty}$ to threshold the importance sampling estimation.
We set the threshold distance at $0.5$. This gave an effective sample size of 275 (out of 1000 samples). This (\textit{i.e.} $\rho=0.5$) may seem large, however it reflects
the change in $G_{y'}$ as $y'$ varies. The shape (and we hope the distortion function $b_y$) of the CDF remains almost unchanged as the location varies.
Data $y'$ with similar $c_{y'}$ functions to $c_y$ are good data so we include as much as we can.
We saw a clear dependence of weight variance on KS-distance. If we set the threshold distance just below 1 (the maximum possible)
the ESS is reduced to 32 as there are some very large weights at larger KS-distances. Data $y'$ at large KS-distance from $y$ is associated with
parameter values $\phi$ that have large IS-weights $1/\tilde p(y|\phi)$, so the KS-distance is helpful in stabilising our estimator in this case.
%We did this for two cases - for $x$ the real data displayed in Figure~\ref{fig:if} and for a second synthetic data
%set in which the data $x$ was a sample, $x\sim p_F(x|\phi)$ on an $M\times M=40\times 40$ square lattice with free boundaries and $\phi=0.3$.
Estimation of $b(y)$ using importance sampling, Algorithm~\ref{al:2} yields $\hat c=0.78$ with standard deviation $\hat \sigma=0.03$, where
\[
\hat \sigma=\sum_{i=1}^M w_i^2(c_i-\hat c)^2.
\]
The convenience of semi-parametric logistic regression in this simple setting is striking. However, importance sampling was also straightforward.

%Suppose an operational coverage of at least 0.75 is deemed acceptable. We can test to see if the operational coverage is worse than 0.75.
%We assume the CLT
%\[
%\frac{\hat c - b(y)}{\hat\sigma}\sim N(0,1)
%\]
%holds to a good approximation. This can be criticised for at least two reasons. First $\hat c$ is consistent for
%$\Pr(\phi\in\tilde C_Y|Y\in \Delta_y)$ not $b(y)$. Also our estimate of $\hat\sigma$ has a much lower ESS (about 32) than the ESS for our estimate
%of $\hat b$ itself (about 275). Since we ran the entire experiment on an inexpensive laptop in few hours without parallelism, this is easily
%remedied by taking $M$ larger and $\rho$ smaller. Assuming the CLT is good and taking a uniform prior for $b(y)\in U[0,1]$, the Bayes factor for
%$H0: b(y)\ge 0.75$ against the alternative $H1: b(y)<0.75$ is
%\[
%B_{0,1}=\frac{\Phi((\hat c-0.75)/0.03)}{1-\Phi((\hat c-0.75)/0.03)}
%\]
%so that $B_{0,1}\simeq 5.3$, which is positive evidence for acceptable coverage.
%the $p$-value in a one-sided test at level $1-a$ with
%$H0: b(y)=0.75$ and alternative $H1: b(y)>0.75$ is $p=1-\Phi((\hat c-0.75)/0.03)$ which gives $p=0.048$. We reject the hypothesis of unacceptably low coverage (subject to the %critical remarks above). The power of this test is

In Figure~\ref{fig:ifr} (right) we plot the estimated calibration function $\hat c_y(\alpha)$. This gives the operational coverage achieved by our estimator
as a function of the nominal coverage we are targeting. This function was estimated as in Algorithm~\ref{al:3} by forming a weighted average of the binary step functions
\[c_i(\alpha)=\mathbb{I}_{\phi_{(i)}\le \tilde q_{y_{(i)}}(\alpha)}\]
defined for $0\le\alpha\le 1$. The black arrow follows the map from a nominal coverage of $0.95$
to the realised coverage (about $0.82$, not a perfect match for the value $0.78$ we estimated using IS in Algorithm~\ref{al:2} due to Monte Carlo error, but note also that
we have switched from equal-tailed to lower-tail credible intervals in moving from Figure~\ref{fig:ifr} (left) to (right)). We can also ask, what nominal level
would give operational coverage equal $0.95$? This is the inverse map represented by the red arrows. We see we should have used $\alpha=0.98$
if we wanted to cover $\phi\sim\pi(\cdot|y)$ $95\%$ of the time.

\section{Mixture-model parameters via Variational Bayes}\label{sec:mix}

Consider data from a mixture of two normal distributions
\begin{equation}\label{eq:mixom}
y\sim p\mathcal N(\mu_1, \sigma_1^2) + (1-p)\mathcal N(\mu_2, \sigma_2^2).
\end{equation}
%In our simulations, $\sigma_1=\sigma_2=1$.
We impose $0<p<\frac12$ to ensure identifiability, and wish to estimate the location of the secondary mode $\mu_1$. To this end, we use Variational Bayes (VB, \citep{jordan1999introduction}). VB provides an analytical approximation to the posterior distribution $\pi(\cdot|y)$, by finding the parametric distribution which minimizes the Kullback-Leibler divergence
\begin{equation}\label{eq:mixkl}
\tilde \pi=\arg\min_{Q\in\mathcal Q}D_{KL}\left(Q(\cdot)||\pi(\cdot|y)\right)
\end{equation}
where $\mathcal Q$ is a parametrized set of distributions. In our example, the set $\mathcal Q$ is defined by imposing that the approximate posterior be of the form
\[(\mu_1,\mu_2, \sigma_1^2, \sigma_2^2, p)\sim \mathcal N(\nu, \tau)\otimes \mathcal N(\nu',\tau')\otimes \mathcal{IG}(a,b)\otimes\mathcal{IG}(a',b')\otimes\mathcal B(a'',b'') \]
for some values of the scalars $(\nu,\tau,\nu',\tau',a,b,a',b',a'',b'')$, where $\mathcal N$, $\mathcal{IG}$ and $\mathcal B$ refer to the Normal, Inverse-Gamma and Beta distributions respectively. We use $\tilde \pi$ defined by Equation~\ref{eq:mixkl} as an approximate posterior distribution.
The computation of the optimal $\tilde \pi$ can be done very rapidly; we used the implementation of the R package vabayelMix \citep{teschendorff2006vabayelmix}.
The R code generating all results in this section is available in the online supplementary material \cite{LNRSupp19}.
We refer the reader to \cite{blei2017variational} for a review of VB, including its application to a mixture of normals.

We implemented Algorithm~\ref{al:1} with $M=10000$ synthetic data sets $y_{(1)},\ldots, y_{(M)}$. For $i=1,\ldots,M$, each data set $y_{(i)}$ is a set of size $n=20$ simulated from the mixture model in Equation~\ref{eq:mixom} with parameters drawn from the priors $\mu_1^{(i)}\sim \mathcal N(0, 10)$, $\mu_2^{(i)}\sim \mathcal N(0, 10)$, $p^{(i)}\sim\mathcal U([0,\frac12])$, and $\sigma_1^{(i)}=\sigma_2^{(i)}=1$. For each synthetic data set $y_{(i)} $, we compute the VB approximate posterior, which we summarize by the set of statistics $s(y_{(i)})=(|\hat\mu_1^{(i)}-\hat\mu_2^{(i)}|, \hat p^{(i)}, {\hat\sigma_1^{(i)}}, {\hat\sigma_2^{(i)}}, \frac{1}{{\hat\sigma_1^{(i)}}}, \frac{1}{{\hat\sigma_2^{(i)}}})$, where $\hat\mu_1^{(i)}$ is the expected value of $\mu_1$ under the VB approximate posterior $\tilde \pi(\cdot|y_{(i)})$, and similarly for the other parameters. This is inspired by the ABC-optimal choice of \cite{fearnhead12}; as with any ABC-like method, the choice of the summary statistics is crucial and including better summary statistics if available can vastly improve the inference: we also experimented with other summary statistics, including the data mean, standard deviation and various quantiles, but found that these statistics did not improve our estimates. We use the VB approximate marginal posterior for $\mu_1$ to compute analytically a 90\% credible interval $\hat C_{(i)}=\tilde C_{y_{(i)}}$ for $\mu_1|y_{(i)}$ and record the binary value $c_i=\mathbb I_{\mu_1^{(i)}\in\hat C_{(i)}}$.

We regressed (using a GAM as above) the coverage indicator $c_i$ against the set of summary statistics $s(y_{(i)})$. Note that these do not form a sufficient statistic, so we should expect some loss of precision. Note also that once this regression is performed, it can be used (at no further computational cost) to estimate the coverage of different observed data sets for the same model.

The output of the regression allows us to estimate the coverage of the VB approximate posterior given the output of VB for some observed data $y$. To evaluate the methodology, we estimated the coverage by simulating $N_{test}=2000$ new ``observed'' data sets. For each data set $j=1\ldots N_{test}$, we used VB to compute an approximate HPD\footnote{Due to the nature of the VB approximation, an HPD interval is an equal-tailed credible interval.} interval $\hat C_j$ with nominal coverage $\alpha=0.90$. We recorded the estimated coverage $\hat c_j\in [0,1]$ given by  Algorithm~\ref{al:1} as well as the binary value $c_j=\mathbb I_{\mu_1^{(j)}\in\hat C_j}$.  We then computed the cross-entropy
\[\frac{1}{N_{test}}\sum_j -c_j\log(\hat c_j) -(1-c_j)\log(1-\hat c_j).\]

%If our estimator is unbiased, then we should have $c_j\sim {\rm Bernoulli}(\hat c_j)$, thus an unbiased estimator would minimize the expected value of the cross-entropy. More generally,
A lower cross-entropy means we are better at estimating the operational coverage. We compare these estimated coverages to the nominal coverage $\forall j, \hat c_j'=0.90$ and to the best constant estimator $\forall j, \hat c_j''=\frac{1}{N_{test}}\sum c_j = 0.719$. Algorithm~\ref{al:1} gave cross-entropy 0.435, $\hat c_j'=0.90$ gave 0.773, and
$\hat c_j''=0.719$ gave 0.616. The fact that we outperform $\hat c_j'=0.90$ indicates that Algorithm~\ref{al:1} estimates the operational coverage better than the nominal coverage. The fact that we outperform $\hat c_j''=0.719$ indicates that we are able to detect in which parts of the space the coverage is higher or lower than average. This kind of experiment is available in general.

\begin{figure}
\includegraphics[width=10cm]{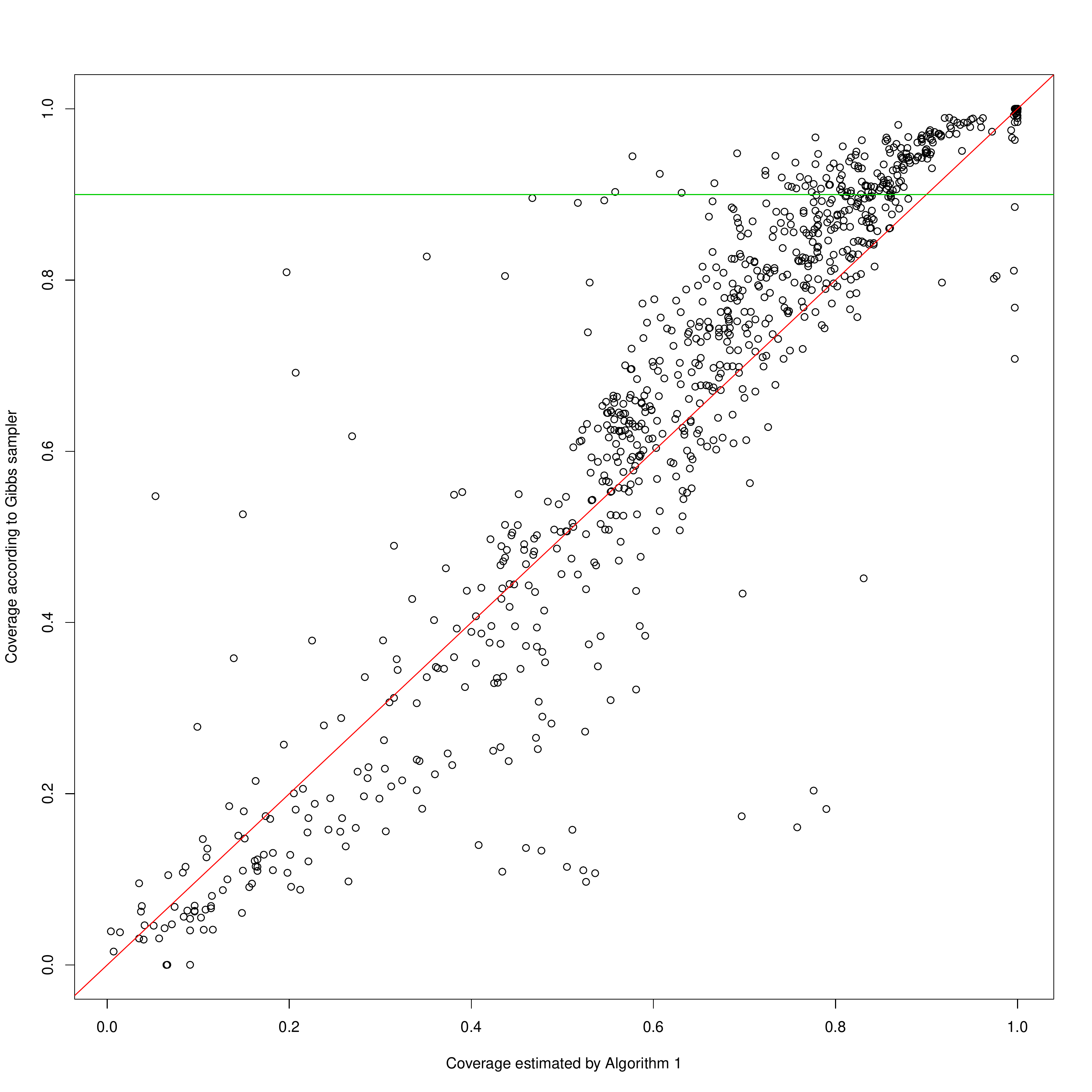}
\caption{\label{fig:Gibbs}
For the mixture of normals example in Section \ref{sec:mix} and $j=1\ldots N_{test}$, true operational coverage $\bar c_j\simeq b(y_{(j)})$ is plotted against estimated operational coverage $\hat c_j$ estimated by Algorithm~\protect\ref{al:1}. Most points are close to the $y=x$ red line. The green horizontal line at $y=0.9$ shows the nominal coverage, which is far from the operational coverage in most cases.}
\end{figure}

For this model, it is also possible to implement Algorithm~\ref{al:0}; we used the Gibbs sampler implementation of the R function \texttt{rnmixGibbs} of package \texttt{bayesm} \citep{rossi2010bayesm} to target $\pi(\phi|y)$. This gives a consistent estimator of the operational coverage of our estimator which would not be available in a real application. We treat this estimate as exact, as we ensured the MCMC sample size was large enough to make the Monte Carlo error negligible.
In order to form this estimate, we generate an MCMC sample $(\theta_{j,1},\ldots,\theta_{j,K})\sim\pi(\cdot|y_{(j)})$. Convergence diagnostics show that $K=1000$ is reasonable. We then record as the true operational coverage the value $\bar c_j=\frac1K\sum_{k=1}^K \mathbb I_{\theta_{j,k}\in\hat C_j}$. The results are shown in Figure \ref{fig:Gibbs}, which plots this ``true'' operational coverage $\bar c_j$ against the estimate $\hat c_j$ given by Algorithm~\ref{al:1}. The vast majority of points are close to the line $y=x$, indicating that our estimator is reliable. We compute the mean squared error $\frac{1}{N_{test}}\sum (\hat c_j-\bar c_j)^2$ (\textit{i.e.} ignoring the small Monte Carlo error in $\bar c_j$). Using the nominal coverage of $0.9$ leads to mean squared error of 0.109; Algorithm~\ref{al:1} leads to a 10-fold improvement with a mean squared error of 0.0106.

\begin{figure}
\includegraphics[width=10cm]{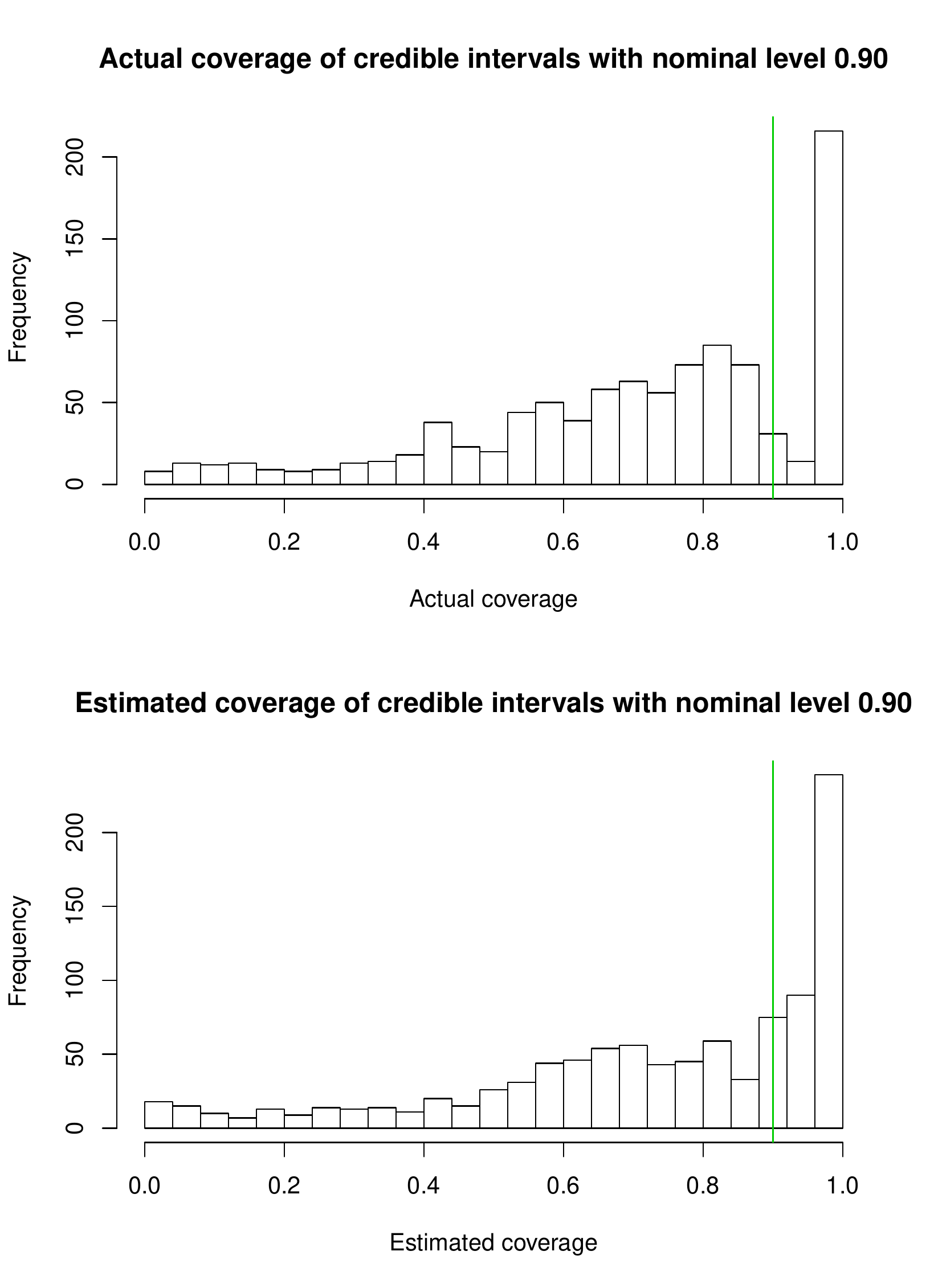}
\caption{\label{fig:VBhist} True and estimated distributions of VB coverage, $x-$ (lower) and $y-$ (upper) marginals of \protect Figure~\ref{fig:Gibbs}.
For 1000 data sets we estimate the operational coverage of VB at that data set using Algorithm~\protect\ref{al:1} (lower, our estimate)
and again using Algorithm~\ref{al:0} and an exact Gibbs sampler (upper, the true operational coverage).
The top plot is a histogram of the actual coverage, which is often very far from the nominal level of 0.90 (green vertical line). The bottom plot shows a histogram of the estimated coverage using our method. Our method correctly identifies the tendency for VB to give low coverage.}
\end{figure}

The mean squared error is a more convincing criterion, but we need an independent consistent estimate of the operational coverage to estimate it, while the cross-entropy can be computed when this is unavailable. For unimodal posterior distributions, VB often provides a good estimate of the posterior mean but underestimates the posterior variance \citep{blei2017variational}, \textit{i.e.} we expect approximate credible intervals to have operational coverage lower than the nominal coverage. This is exemplified in Figure \ref{fig:VBhist} (top), which gives the actual coverage of 1000 approximate credible intervals each with nominal level 0.90.

\section{Coverage of a partition of random effects}\label{sec:car}

The \texttt{car90} data contain specifications of $n=111$ cars, extracted from {\it Consumer Reports, 1990}. The dependence of car price on car specifications is of interest. The dataset is available in the R package \texttt{rpart} \citep{therneau2015rpart}. We focus on the problem of
clustering the levels of a categorical variable as part of the modeling.
We use these data to illustrate an approximate method for fitting a Dirichlet process model for the clustering. The output of this analysis is a credible set of partitions
of the levels indicating how the levels may plausibly be grouped.
For this purpose we select from the original 33 variables the engine displacement in cubic inches ($x=(x_1,\ldots,x_n)$), the red line value (the maximum safe engine speed in rpm,
$z=(z_1,\ldots,z_n)$) and the car type ($t=(t_1,\ldots,t_n)$ with $t_i\in \mathcal T$ for $i=1,\ldots,n$ and $\mathcal{T}=\{\mbox{small, medium, large, van, compact, sporty}\}$).

Let $S=(S_1,\ldots,S_K)$ be a partition of $\mathcal{T}$ into $K$ sets,
with $K\in \{1,2,\ldots,6\}$, and for $i=1,\ldots,n$ let $s_i$ denote the partition for car $i$ so that $s_i=k$ is equivalent to $t_i\in S_k$. In our model the overall effects due to type are assumed to fall in groups: we have a separate effect $\gamma_k$ for each group $k=1,\ldots,K$, and an effect for each type within each group, $\eta_\tau, \tau\in \mathcal T$.
The two random effect covariance matrices $\Sigma_\eta$ and $\Sigma_\gamma$ are assumed diagonal with diagonal elements $h_\eta\sigma^2$ and $h_\gamma\sigma^2$ respectively,
where $\sigma^2$ is defined below as the response variance.
The overall random effect, $\eta'_{t_i}$ say, for observation $i$ is $\eta'_{t_i}=\eta_{t_i}+\gamma_{s_i}$ (notice $s_i$ is determined from $t_i$ given $S$).
Let $\gamma=(\gamma_1,\ldots,\gamma_K)$ and $\eta=(\eta_1,\ldots,\eta_6)$.

The model in this section clusters random effects ``by covariance''. If we integrate out $\gamma \in \R^K$ given the partition $S$ then we are left with a model in which random effects in the same group have a higher covariance (\textit{i.e.} $h_\gamma\sigma^2$) than random effects in different groups (where the covariance zero). \cite{Walli2018} and \cite{pauger18} treat the same problem in a similar way but use a different parameterisation and prior. \cite{Walli2018} use a mixture of spiky normal distributions in their prior for label effects. \cite{pauger18} take the overall random effects for type $\eta'_\tau, \tau\in \mathcal{T}$ and introduce covariance terms off-diagonal in the random effects covariance matrix $\Sigma_{\eta'}$. They operate directly on the elements of the covariance matrix in order to explore the model space.
%due to the type-group and The effect by car type ($Type$) is modelled in two ways: type specific effect $\eta_{\rm type}$ and partition effect $\gamma_{\rm type,s}$.
%Given six car types, there are six type specific effects. Six car types are grouped with similar effect and types in the same group are expected to have the same effect. A partition of types is denoted by $s$ and the number of partition sets is as small as one or as large as six.

Given the partition $S$ the price $y$ in \$1000 dollars is modelled using the following random and fixed effects, for $i=1,\ldots,n$,
\[ y_i = \beta_0 + \beta_{1} x_i + \beta_{2} z_i + \eta_{t_i} + \gamma_{s_i} + \epsilon_i \]
where $\epsilon_i \sim \N(0,\sigma^2)$. The parameter priors are
\[ \pi(\sigma^2) \propto 1/\sigma^2 \,, \hspace{.3cm} \beta_0,\beta_{1},\beta_{2} \sim \N(0,h_b\sigma^2)\,, \hspace{.3cm}  \eta_{\tau} \sim \N(0,h_{\eta}\sigma^2), \tau\in \mathcal{T} \,.  \]
The partition $S$ is unknown. Let $\mathcal{P}$ denote the space of partitions of our six types (there are 203 distinct partitions). We take a Chinese restaurant process (CRP) prior $\pi(S)$ for $S$, with clustering parameter $\alpha_{CRP}=1$
\[ \pi(S) = \dfrac{\Gamma(\alpha_{CRP})}{\Gamma(\alpha_{CRP}+n)}\alpha_{CRP}^K\prod^K_{k=1} \Gamma(n_k)  \]
where $n_k$ is the number of datapoints in the partition $k$.
 We are modeling the random effects via a Dirichlet Process $G_\gamma\sim DP(\alpha,H), \gamma_{k} \sim G_\gamma$
 with base density
\[ H(\gamma_k)=\N(\gamma_k;0,h_{\gamma}\sigma^2), k=1,\ldots,K  \,. \]
Integrating over the DP random measure $G_\gamma$ yields the prior
\[
\pi(\gamma,S|h_\gamma,\sigma^2)=\pi(S)\prod_{k=1}^{K(S)}\N(\gamma_k; 0,h_{\gamma}\sigma^2).
\]
The scale parameter priors are $h_\eta,h_\gamma,h_b \sim Inv\chi^2(1)$. Let $h=(h_\eta,h_\gamma,h_b)$.

We are interested in the marginal posterior distribution of $S|y$ and estimating a HPD credible set for $S$. Let $$\psi=(\beta_0,\beta_{1},\beta_{2},\sigma,\gamma,\eta)$$ denote the vector of parameters besides $S$ and $h$. It is often convenient (\textit{i.e.} in models slightly more complex that this) to work directly with the marginal (or ``collapsed'') posterior $$\pi(S|y)\propto p(y|S)\pi(S),$$ where $p(y|S)$ is the intractable marginal likelihood
\[
p(y|S)=\int p(y|S,h,\psi)\pi(\psi|h,S)\pi(S,h)d\psi dh.
\]
However $p(y|S,h)$ is available in closed form and there are a number of ways one might then proceed to solve the problem without further approximation (for example using asymptotically exact MCMC). Here we simply set
$h_\eta=h_\gamma=h_b=10$, that is we define
\[
\tilde p(y|S)=\left.p(y|S,h)\right\vert_{h_\eta=h_\gamma=h_b=10}
\]
and
\[
\tilde\pi(S|y)\propto \tilde p(y|S)\pi(S).
\]

We implemented MCMC targeting $\tilde\pi(S|y)$ using Metropolis Hastings MCMC updating one level of car type at each update.
We define and estimate the empirical CDF $\hat G_{y'}(S),\ s\in \mathcal P, y'\in\mathcal Y$ and associated KS distance as follows. For $S\in \mathcal P$ let
$\hat \pi(S|y)$ be an estimate of the approximate posterior formed from the MCMC output, and let $>_{y}$ be the (random) order on partitions
determined by $S >_y S'\Leftrightarrow \hat\pi(S|y)>\hat\pi(S'|y)$. The empirical cdf of the approximate posterior at $y'\in\mathcal Y$ is
$\hat G_{y'}(S)=\sum_{S'\ge_y S} \hat\pi(S|y')$ and the estimated KS distance is $\delta(y',y)=\left\lVert \hat G_{y}-\hat G_{y'}\right\rVert_{\infty}$.

%approximately in two different ways. In one we estimate $p(y|S)$ by averaging $p(y|S,h_\eta,h_\gamma,h_b)$ using draws from the priors for
%$h_\eta,h_\gamma,h_b$ (we appreciate there are several ways to solve the problem exactly but we )
%The co-occurrence matrix of car types in common partition sets
%is shown in Figure \ref{fig:car}. This is based on 1,000 posterior partition samples.
%There is a clear partition of 'Medium' from the other types and no clear partition among those five types.
%
%\begin{figure}
%	\centering
%	\includegraphics[width=1.8in]{car_coom_M1}
%	\caption{Posterior estimate for the co-occurrence matrix using $\tilde m_h$ with $h_\eta=h_\gamma=h_b=10$}\label{fig:car}
%\end{figure}
The level $\alpha=0.95$ HPD set is shown in Table~\ref{tab:car}. We would like to use our calibration check to see if this Monte Carlo HPD set
is reliable. We estimated the coverage probability of each partition using importance sampling, Algorithm~\ref{al:2}, with $M=100000$ samples and the KS-distance function.
In this multi-parameter setting we need $\psi$ and $h$ in order to simulate $p(y|S,h,\psi)$. These can often be sampled from their priors,
so the only importance re-weighting comes with replacing $S\sim \pi(\cdot)$ with $S\sim \tilde\pi(\cdot|y)$. However here $\sigma$ has an improper
prior, so we take $\phi=(\psi,S)$ in Algorithm~\ref{al:2} and sample $\psi,S|y\sim \tilde\pi(\cdot|y)$, where $\tilde\pi(\psi,S|y)\propto\pi(\psi,S,h|y)\vert_{h=10}$ is the approximate posterior for all the parameters. When we use this importance sampling proposal distribution, the IS weight is $w(S,\psi)\propto 1/\tilde p(y|S,\psi)$.

\begin{figure}
\vspace*{-1in}
	\centering
	\hspace*{-0.1in}\includegraphics[width=2.6in]{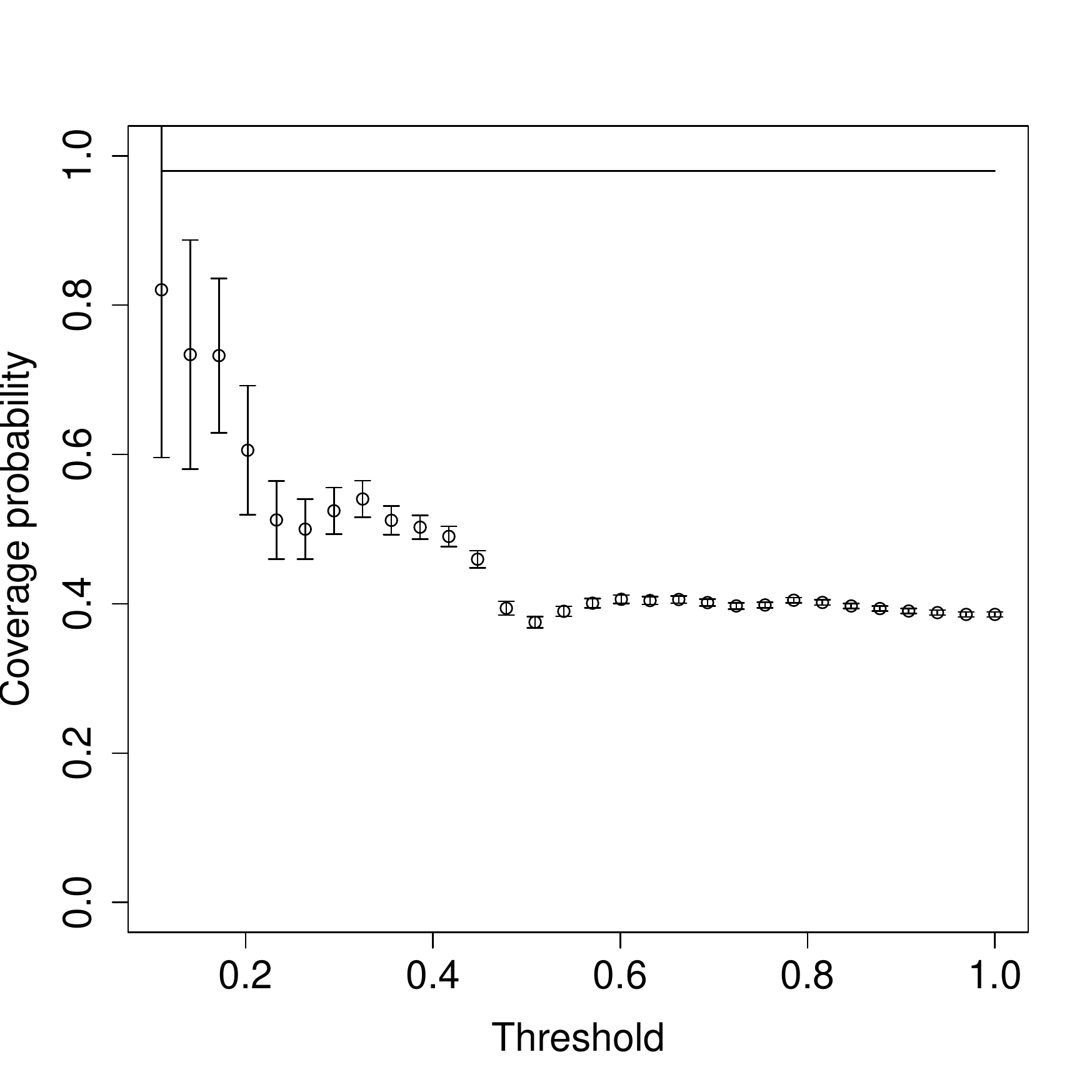}\hspace*{0.3in}\includegraphics[width=2.6in]{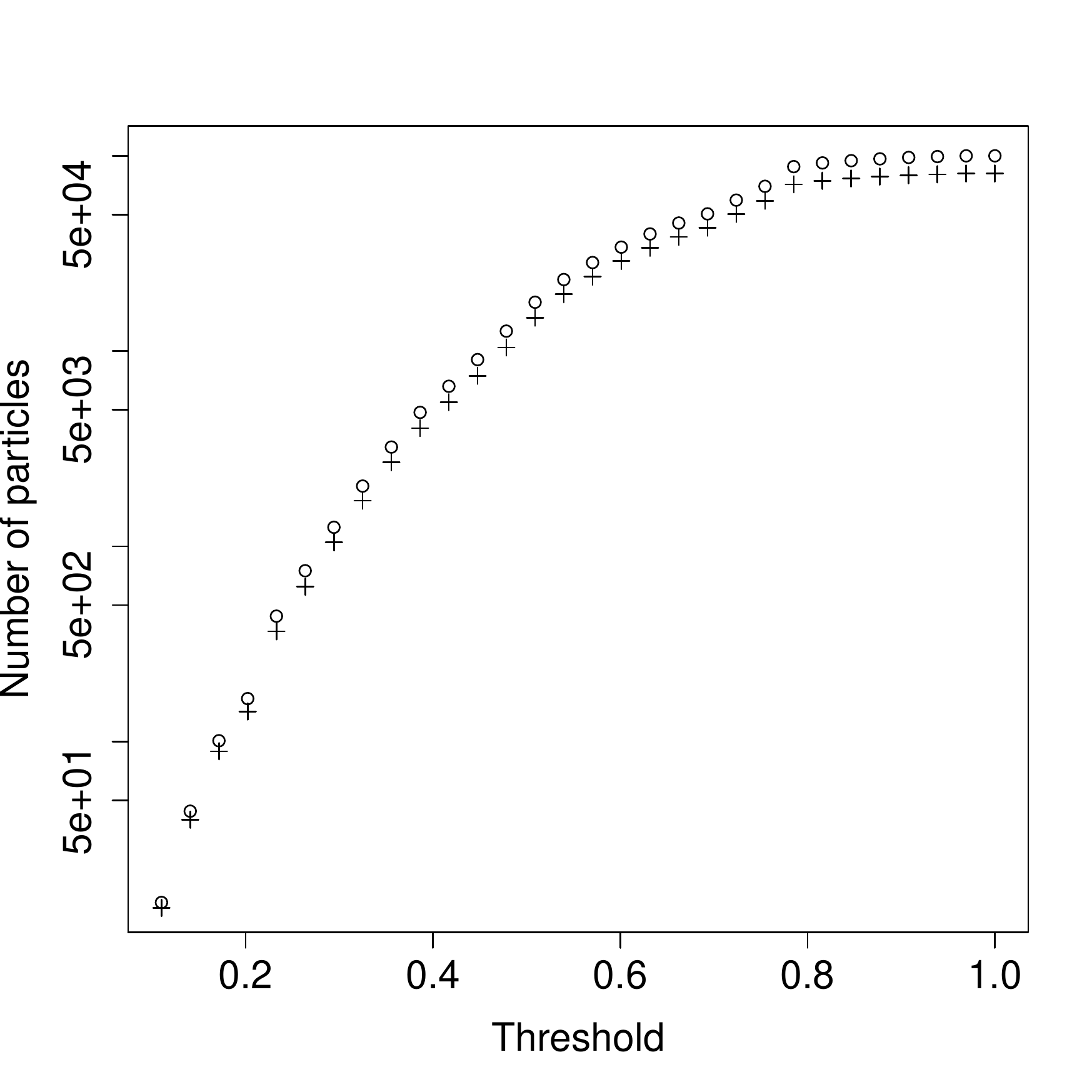}
	\caption{ (Left) Coverage probability estimates $\hat c(y)$ for Section \ref{sec:car}. For each threshold value $\rho$ a KS-distance $\delta(y',y)$ and an operational coverage probability $c(y)$ are estimated. Error bars are 2-sigma. Horizontal line (at $0.98$) is a high precision estimate of $c(y)$ using Algorithm~\protect\ref{al:0}.
    (Right) Effective Sample Sizes (ESS) using Importance sampling. The number of particles in the $\Delta_y$ neighbourhood (circle) and the ESS (cross) of the weighted samples surviving the cut, as a function of $\rho$. }\label{fig:car:coverage}
\end{figure}

%AE1 comment 8 - not KS but CDF of data. Could possibly expand this with math detail and say a few words about symmetry.
%We define the CDF for our distribution over partitions by ordering the partitions according to their probability in $\tilde \pi(S|y)$. The CDF at $S=s$
%for $\tilde \pi(S|y')$ is a cumulative sum along this standard order from the mode to $S=s$.  The KS-distance is then defined on CDF's in the usual way.
%Here we report results for a much simpler distance which seems to work well here, perhaps due to the symmetry of $c(y)$ under permutation of
%covariate levels. % estimates in Figure~\ref{fig:car:coverage}).
%We define the distance between two datasets as the KS-distance between
%the empirical CDF's of the simulated data itself. This has the advantage that it does not require simulation of $\theta|y'$ to determine $\delta(y',y)$.
%%is defined KS-distance for partitions is not straight forward to estimate and, for this study we instead use the empirical CDFs of data ($y$) and simulated data from the prior.
Results are summarised in Figure~\ref{fig:car:coverage}. We additionally estimate the operational coverage (at $c(y)\simeq 0.98$) using Algorithm~\ref{al:0} to get an unbiased and consistent estimate of $c(y)$ with very small uncertainty. We regard this as the truth (horizontal line in graph at left in Figure~\ref{fig:car:coverage}). This would not in general be available (if we can implement Algorithm~\ref{al:0} we can sample $\pi(\phi|y)$). The coverage probability estimate varies sharply with $\rho$ and
approaches the nominal coverage at $\alpha=0.95$ at small $\rho$ with an ESS dropping from 50 to 15 in the last two (leftmost)
point estimates. We see that $c(y)$ varies significantly with $y\in\mathcal Y$ but the coverage is likely to be good. This is a hard estimation problem,
and exposes the need for better estimators.

%\begin{table}
%	\begin{tabular}{c|c}
%		Partition, $s$ & $G(s|Price)$ \\ \hline
%        (Compact,Large,Medium,Small,Sporty,Van) &0.68 \\
%      (Compact,Large,Medium,Sporty,Van),(Small) &0.72 \\
%      (Compact,Large,Medium,Small,Sporty),(Van) &0.75 \\
%      (Compact,Large,Small,Sporty,Van),(Medium) &0.78 \\
%      (Compact,Medium),(Large,Small,Sporty,Van) &0.80\\ \vdots & \vdots \\
%		  (Compact,Small),(Large,Sporty,Van),(Medium) 0.9490 & 0.95 \\
%		\end{tabular} \caption{HPD set for partition $s$ using $\tilde m_h$ with $h_\eta=h_\gamma=h_b=10$. Columns indicate partitions sorted by $P(s|Price)$ and cumulative sum (ie, CDF) $G(s|Price)$.}\label{tab:car}
%\end{table}

\begin{table}
	\begin{tabular}{c|c}
		Partition, $S$ & $G(S|y)$ \\ \hline
           (Compact,Large,Medium,Small,Sporty,Van) & 0.21 \\
         (Compact,Large,Medium,Sporty,Van),(Small) & 0.25 \\
         (Compact,Medium,Small,Sporty,Van),(Large) & 0.29 \\
         (Compact,Large,Medium,Small,Sporty),(Van) & 0.32 \\
         (Compact,Large,Medium,Small,Van),(Sporty) & 0.36 \\ \vdots & \vdots \\
		  (Compact,Van),(Large,Sporty),(Medium,Small) & 0.95 \\
		\end{tabular} \caption{HPD set for partition $S$ in Section \ref{sec:car} using $\tilde \pi(S|y)$. Rows are partitions sorted by posterior probability with the largest first.
        The second column is the cumulative sum (\textit{i.e.} the CDF $\hat G_y(s)$). There are 144 partitions in this HPD set.}\label{tab:car}
\end{table}

%\begin{figure}
%	\centering
%	\includegraphics[width=7in,height=2in]{car_coverage_prob_1}
%	\includegraphics[width=7in,height=2in]{car_coverage_prob_2}
%	\includegraphics[width=7in,height=2in]{car_coverage_prob_3}
%	\caption{Three marginal likelihood approximations are $\tilde m_h$ with $h_\eta=h_\gamma=h_b=10$ (Top row), $\tilde m_h$ with $h_\eta=h_\gamma=h_b=1$ (Middle row) and $\tilde m_{\rm BIC}$ (Bottom row). For the distance measure $d(y',y)$, the Euclidean distance (First and Second columns) and the KS distance (Third and Fourth columns) are used. For each approximation and for each threshold value of $d(y',y)$, a coverage probability is estimated using the importance sampling (First and Third column) and, a number of particles in the neighbourhood is marked by circle and an effective sample size, marked by cross (Second and Fourth column). }\label{fig:car:coverage}
%\end{figure}

\section{Conclusions}

%If diagnostics are correct then marginal expectations are correct.

We have presented two methods for estimating the operational coverage of approximate credible sets. Our examples show that the operational coverage can be very far from the nominal coverage; the nominal coverage should thus not be taken at face value. When we are in control of the precision of our approximation it may be convincing simply to check that credible sets are stable as precision is increased. However when we make a fixed approximation, as we do in Sections~\ref{sec:part}, \ref{sec:mix} and~\ref{sec:car} this standard check is no longer available, and in all cases a measure of the operational posterior coverage our posterior approximation achieves will be of interest.

%One approach we considered is to sample $\phi\sim\tilde\pi(\cdot|y), y'\sim p(\cdot|\phi)$ and $\theta\sim\tilde\pi(\cdot|y')$
%and report the proportion of $\hat C_{y'}(\theta)$-sets covering $\phi$.
%%keeping $(\phi,y',\theta)$ satisfying $y'\in\Delta_y$.
%Notice that the data $y'$ is simulated under the true model but analysed under an approximation.
%This is Algorithm~\ref{al:2} but without windowing on the data or IS-reweighting. This approach makes sense if we think the unknown true parameter $\phi$
%really is a typical state in the distribution determined by $\tilde\pi(\phi|y)$. Since this procedure averages over $\phi$ and $y'$,
%the coverage it estimates includes variation due to data realisation for $\phi$ values we believe are
%close to the true parameter. It is easily estimated.
%However it does not give the nominal level when there is no approximation, and if the approximation is poor,
%we may be measuring coverage in the wrong part of $(\phi,y)$-space.

Depending on the setting, Algorithm~\ref{al:1} or~\ref{al:2} may be easier to implement. We have found that both algorithms are relatively straightforward to implement, and both can apply to a wide range of approximation algorithms. The principal weakness of our approach is judging the reliability of our reliability checks. When our coverage estimators failed %the failure was spectacular (estimated coverage equal zero or one) and
the reason for the failure was obvious (very small ESS or unreliable extrapolation in high dimension).
A coverage estimate conditioned on the data and based on Equation~\ref{eq:iec} can still be fooled, but in ways that differ from tests averaging over data, and based on Equation~\ref{eq:iem}. However more subtle errors will often be picked up using standard checks on stability. In the example in Section~\ref{sec:part}
standard logistic regression checks provide good evidence that our reliability estimate is itself reliable. In more complex settings it may be necessary to find estimators with lower mean square error. The estimators we give are simple and could be improved a great deal.

Approximate inference schemes are essential tools in Bayesian analysis, and are gaining in popularity. We suggest that estimating the calibration of approximate credible sets can vastly improve the trust we have in the output of such schemes, or alternatively serve as a check when the approximations fail, and that the non-negligible computational time of these methods is a fair price to pay for such a check.

\section*{Appendix A: IS weight variance}

In this appendix we prove that the estimator $\hat d$ for $d=d(y)$ given in Section~\ref{sec:ne} satisfies a CLT, that is
\begin{equation}\label{eq:clt}
\sqrt{M}(\hat d-d)\rightarrow \N(0,V),
\end{equation}
with $V$ finite. We keep the presentation general as far as we can.

We wish to estimate $d(y)$ where
\[
d(y)=\int_{\Omega}\int_{\mathcal Y} \mathbb{I}_{\phi\in \tilde C_{y'}} m(\phi,y') dy' d\phi,
\]
and
\[
m(\phi,y') = {z(y)}^{-1} \pi(\phi)p(y'|\phi)\mathbb{I}_{y'\in \Delta_y},
\]
a joint density with normalising constant $z(y)$. The $\theta$-integration has disappeared from the definition of $d(y)$ given in Equation~\ref{eq:ist2}, because
$\hat d$ in Section~\ref{sec:ne} is given in terms of credible intervals $\tilde C_{y'}$ which are exact for $\tilde\pi(\theta|y')$ and not estimates $\hat C_{y'}$ based on sampled $\theta$ (ie we want the operational coverage $b(y)$ rather than the realised coverage $c(y)$).
Let $f(\phi,y')=\mathbb{I}_{\phi\in \tilde C_{y'}}$ so that $d(y)=E_m(f(\phi,y'))$.
Let $\tilde m(\phi,y')$ be the joint IS density with normalising constant $\tilde z(y)$,
\[\tilde m(\phi,y') = {\tilde z(y)}^{-1} \tilde \pi(\phi|y)p(y'|\phi)\mathbb{I}_{y'\in \Delta_y}.\]
Since $\tilde \pi(\phi|y)=\pi(\phi)\tilde p(y|\phi)/\tilde p(y)$, the {\it normalised} importance weight function
$m/\tilde m$ is $w(\phi;y) = k(y)/\tilde p(y|\phi)$ with $k(y)=\tilde p(y)\tilde z(y)/z(y)$. We now drop the explicit dependence on $y$ from our notation as $y$ is fixed.
The {\it unnormalised} weight function is $\tilde w(\phi) = 1/\tilde p(y|\phi)$, so the unnormalised importance weights
are for $i=1,...,M$, $\tilde w_i=\tilde w(\phi_{(i)})$ and our IS estimator is $\hat d=\sum_i \tilde w_i c_i/\sum_j \tilde w_j$, with $c_i=f(\phi_{(i)},y_{(i)})$
and $(\phi_{(i)},y_{(i)})\sim \tilde m$ for $i=1,...,M$.

%A posterior $\pi(\phi|y'), \phi\in\Omega$ is a proper probability distribution at $y'$
%if it integrates to one over $\phi\in \Omega$.
If the posterior $\pi(\phi|y')=\pi(\phi)p(y'|\phi)/p(y')$
is proper then $p(y')$ is finite at $y'$.
Since $\pi(\phi|y')$ in Section~\ref{sec:ne} is a normal density with finite non-zero variance and mean $y'/2$, it is proper at every $y'\in \mathcal{Y}$.
The same applies to $\tilde\pi(\theta|y')$ in Equation~\ref{eq:nap}, so $p(y')$ and $\tilde p(y')$
are finite at each $y'\in \mathcal{Y}$.
Because $\pi(\phi|y')$ and $\tilde\pi(\theta|y')$ are proper posteriors, it follows that
$m$ and $\tilde m$ are proper densities. Also, $z$, $\tilde z$ and $k$ are finite, $E_{\tilde m}(w(\phi))=1$ and $d=d(y)$ itself is finite.

%This is the self normalised IS estimator, estimating the quantity $d(y)$ defined in Equation~\ref{eq:ist2}.
%This corresponds to the setting described in Section 4.3 of \citet{robert2009introducing}, which follows \citet{liu1996metropolized} and \citet{robert2004monte}. Letting $S_f^n = \sum \tilde w_i f(\phi_{(i)},y_{(i)})$ and $S_1^n = \sum \tilde w_i$, they show that
%\[\var(\hat d) = \frac{1}{n^2 k^2}\left( \var_{\tilde m} (S_h^n) - 2 E_m[f] cov_{\tilde m}(S_h^n, S_1^n) + E_m[f^2]\var_{\tilde m} (S_1^n)\right).\]
%Since $f$ takes values in $\{0, 1\}$, this variance can be bounded by
%\[\var(\hat d) \leq \frac{4}{n^2 k^2} \var S_1^n \leq \frac{4}{n k^2} E_{\tilde m}[w^2].\]
%Note that $E_{\tilde m}[w^2] = E_m[w]$; $\hat d$ has finite variance if this quantity is finite.

We now show that $\hat d$ has a CLT using the Multivariate Delta-Method. %The statement in \cite{casella2002statistical}
%is convenient for us, as it  CLT with the Delta-Method in one result.
%The calculations which follow are routine, once we know the Delta-Method applies.
%We show that it does apply, without assuming the moments  are finite.
Let $\bar d=M^{-1}\sum_i \tilde w_i k c_i$ and
$\bar w=M^{-1}\sum_i k \tilde w_i$ denote {\it normalised} IS estimates, so that
$\hat d=\bar d/\bar w$. Let $g(a,b)=a/b$, so that $\hat d=g(\bar d,\bar w)$,
with $g_a=1/b$ and $g_b=-a/b^2$ partial derivatives.
Let $\var(\bar d)=U_d/M$, $\var(\bar w)=U_w/M$ and $\cov(\bar d,\bar w)=U_{d,w}/M$.
If these quantities are all finite then by Theorem 5.5.28 of \cite{casella2002statistical},
\[\sqrt{M}(g(\bar d,\bar w)-g(d,1))\rightarrow \N(0,V)\]
as $M\rightarrow\infty$, with
\begin{eqnarray}
% \nonumber to remove numbering (before each equation)
  V &=& (U_d g_a^2 + 2U_{d,w}g_a g_b + U_w g_b^2)\vert_{(a,b)=(d,1)} \nonumber \\
    &=& U_d - 2dU_{d,w} + d^2 U_w. \label{eq:V}
\end{eqnarray}
%\begin{eqnarray*}
% \nonumber to remove numbering (before each equation)
%  \var(\hat d) &=& \var\left(\sum_i \tilde w_i c_i/\sum_j \tilde w_j\right) \\
%    %&=& \var\left(\sum_i \tilde w_i k c_i/\sum_j k \tilde w_j\right)\\
%    &=& \var(\bar d/\bar w),
%\end{eqnarray*}
%where we have multiplied by $k$ top and bottom, so in the final line $\bar d=M^{-1}\sum_i \tilde w_i k c_i$ and
%$\bar w=M^{-1}\sum_j k \tilde w_j$ are normalised IS estimates.
Now since $E_{\tilde m}(\bar d)=E_{\tilde m}(f(\phi,y')w(\phi))$ is equal $d$, we have
\begin{eqnarray*}
% \nonumber to remove numbering (before each equation)
  U_d&=& E_{\tilde m}\left((f(\phi,y')w(\phi)-d)^2\right) \\
   &=& E_m(f(\phi,y')w(\phi))-d^2,
\end{eqnarray*}
expanding the top line and using $E_{\tilde m}(fw^2)=E_m(fw)$ and $f^2=f$.
The expectation of $\bar w$ is $1$. Also, $U_w=E_m(w(\phi))-1$ (set $f, d$ equal $1$ in $U_d$)
%Finally the covariance of $\bar d$ and $\bar w$, $\cov(\bar d,\bar w)=U_{d,w}/M$ is given by
and
\begin{eqnarray*}
% \nonumber to remove numbering (before each equation)
  U_{d,w} &=& E_{\tilde m}\left(\bar d\bar w\right)- d \\
   &=&E_m(f(\phi,y')w(\phi))-d.
\end{eqnarray*}
These quantities are all finite if $E_m(w(\phi))$ is finite because $f\le 1$ so $E_m(f(\phi,y')w(\phi))\le E_m(w(\phi))$.
In that case $V$ in Equation~\ref{eq:V} is finite\footnote{Substituting for $U_d, U_w$ and $U_{d,w}$ and using $E_{\tilde m}(fw^2)=E_m(fw)$ again
gives
$
V=E_{\tilde m}\left(w(\phi)^2(f(\phi,y')-d(y))^2\right)
$
the standard result (see for eg \cite{owen13})} and we are done.

%, then $M\var(\hat d)\rightarrow V$ by the Multivariate Delta-Method and so the variance of $\hat d$ must be finite for all sufficiently large $M$.
%
%
%
%
%
%Now $\bar d$ is the normalised IS estimate for $d$ with variance $U_M/M$ where
%\[
%U_M = E_{\tilde m}\left((f(\phi,y')w(\phi)-d(y))^2\right).
%\]
%If $U_M$ is finite then $\bar d$ satisfies the CLT $\sqrt{M}(\bar d-d)\rightarrow \N(0,U_M)$ and $\bar w\rightarrow 1$
%as $M\rightarrow \infty$. In this case the conditions for the Delta-method to apply are satisfied {\color{blue}(CHECK THIS)}.
%The Delta method gives $\var(\hat d)\simeq V_M/M$ where (\cite{owen13})
%\[
%V_M = E_{\tilde m}\left(w(\phi)^2(f(\phi,y')-d(y))^2\right).%\\
%\]
%If $V_M$ is finite then $M\var(\hat d)/V_M\rightarrow 1$ as $M\rightarrow\infty$ {\color{blue}(CHECK THIS)}.
%It follows that if $U_M$ and $V_M$ are finite then $\var(\hat d)$ is finite.
%%so $\sqrt{M}(\bar d/\bar w-d)=\sqrt{M}(\bar d - d)/\hat w^*+\sqrt{M}(1/\hat w^*-1)$.
%%to show $\hat d^*/\hat w^*\rightarrow \N(d,V_M)$, and hence $\var(\hat d)/V_M\rightarrow 1$.
%Expand $U_M$ and $V_M$ and use $E_{\tilde m}(fw^2)=E_m(fw)$ and $f^2=f$ to get
%\[
%U_M = E_m(f(\phi,y')w(\phi))-d(y)^2
%\]
%and
%\[
%V_M = E_m(f(\phi,y')w(\phi))-2d(y)E_m(f(\phi,y')w(\phi))+d(y)^2E_m(w(\phi)).
%\]
%Now since $E_m(f(\phi,y')w(\phi))\le E_m(w(\phi))$ it follows that $E_m(w(\phi))<\infty$
%is sufficient for both $U_M$ and $V_M$ to be finite.

In our setting, with $k$ and $z$ finite constants not depending on $y'$ or $\phi$,
\begin{eqnarray*}
% \nonumber to remove numbering (before each equation)
  E_m(w(\phi)) &=& \int_{\Omega}\int_{\mathcal{Y}} \frac{k}{p(y|\phi)} m(\phi,y')\,dy' d\phi \\
    &=& \frac{k}{z}\int_{\Delta_y} \int_{\Omega}\pi(\phi)\frac{p(y'|\phi)}{\tilde p(y|\phi)}\,d\phi dy'.
%    &=& \frac{k}{z} \int_{\Delta_y} h(y') dy'
\end{eqnarray*}
%If the posterior $\tilde \pi(\theta|y')$ is proper, as here, and $\Delta_y$ is closed and bounded (which it is for the Euclidean distance used in Section~\ref{sec:ne})
%then $\tilde p(y')$ is bounded in $\Delta_y$ (from the discussion above) so
%for $\var(\hat d)$ to be finite it is sufficient to show that the expectation
Since $k/z$ is finite, $E_m(w(\phi))$ is finite if $h(y';y)=E_\pi(p(y'|\phi)/\tilde p(y|\phi))$
%\[
%E\left(\frac{p(y'|\phi)}{\tilde p(y|\phi)}{\left\vert\right.} y'\in \Delta_y \right)=
%h(y')=\int_{\Omega}\pi(\phi)\frac{p(y'|\phi)}{\tilde p(y|\phi)}d\phi
%\]
is bounded on the compact set $y'\in \Delta_y$.
%Now $p(y'|\phi)/\tilde p(y|\phi)$ is not in general bounded for $y'\in \Delta_y$ and all $\phi$.
%In the case above we have $\tilde p(y'|\phi)=p(y'|\phi)$ for $y'\in \Delta_y\cap [-3,3]$, with
%\[
%Z(\phi)=\Phi(-3.5-\phi)+\Phi(3-\phi)-\Phi(-3-\phi)+1-\Phi(2.5-\phi),
%\]
%and $Z(\phi)\le 3$ (no attempt at a tight bound here) so
%$E_{y';y}$ is finite (and less than $3$) in that case.
For the densities we have in Section~\ref{sec:ne}
\begin{eqnarray*}
h(y';y)&\propto&\int_{-\infty}^\infty \N(\phi;0,1)\exp\left(-\frac{(\phi-y')^2}{2}\right)\exp\left(\frac{v(\phi-y)^2}{2}\right)\,d\phi \\
&\propto&\int_{-\infty}^\infty \exp\left(-\frac{\phi^2(2-v)}{2}+\phi (y'-vy)\right)\,d\phi,
\end{eqnarray*}
which is finite for $0\le v<2$ and any fixed $y'$.

%AE1 comment Appendix/3. Inserted explanation
We draw some general lessons. The IS estimator $\hat d$ for operational coverage $b(y)$ 
will have a CLT if the posteriors are proper and $E_\pi(p(y'|\phi)/\tilde p(y|\phi))$ is bounded for 
$y'\in \Delta_y$. As usual in IS, problems arise when $\phi$-variation in the IS proposal (here the approximate posterior) is under-dispersed with respect
to the IS target (here exact posterior), 
as this leads to large weight values. In our example things work well when $v$ is small ($ie$ $0\le v<2$): when $0\le v<1$ the approximation
is over-dispersed with respect to the exact posterior. When $v>1$ it is under-dispersed. There is some margin above $v=1$ due to the damping effect of the prior.
If the approximation is under-dispersed then we might try Algorithm~\ref{al:1} (regression) ahead of Algorithm~\ref{al:2} (IS).

%\section*{Appendix B: Code}
%The code for the computation of the normalizing constant in Section \ref{sec:part}, as well as some code to reproduce the results of Section \ref{sec:mix}, are available on github at \url{https://github.com/robinryder/calibration}.

\bibliographystyle{ba}
\bibliography{reference}

\end{document}